\documentclass[useAMS,twocolumn,usenatbib]{mnras}
\pdfoutput=1
\usepackage{times,psfig,epsfig}
\usepackage{rotating, graphicx}
\usepackage{color}
\usepackage{amssymb, amsmath}
\usepackage[normalem]{ulem}
\usepackage[titletoc]{appendix}
\usepackage{adjustbox}
\usepackage{booktabs}
\def\aap{A\&A}

\def\apjl{ApJL}

\def\apjs{ApJS}

\newcommand{\bdm}{\begin{displaymath}}
\newcommand{\edm}{\end{displaymath}}
\newcommand{\beq}{\begin{equation}}
\newcommand{\eeq}{\end{equation}}
\newcommand{\beqnarr}{\begin{eqnarray}}
\newcommand{\eeqnarr}{\end{eqnarray}}
\newcommand{\bit}{\begin{itemize}}
\newcommand{\eit}{\end{itemize}}
\newcommand{\ben}{\begin{enumerate}}
\newcommand{\een}{\end{enumerate}}
\newcommand{\bfi}{\begin{figure}[htb]}
\newcommand{\bpfi}{\begin{figure}[p]}
\newcommand{\barr}{\begin{array}}
\newcommand{\earr}{\end{array}}
\newcommand{\bec}{\begin{center}}
\newcommand{\eec}{\end{center}}
\newcommand{\bs}{\begin{sideways}}
\newcommand{\es}{\end{sideways}}

\newcommand{\ATLAS}{\textrm{ATLAS}^{\rm 3D}}

 \newcommand{\mincir}{\raise
  -2.truept\hbox{\rlap{\hbox{$\sim$}}\raise5.truept \hbox{$<$}\ }}
\newcommand{\magcir}{\raise
  -2.truept\hbox{\rlap{\hbox{$\sim$}}\raise5.truept \hbox{$>$}\ }}
\newcommand{\siml}{\raise
  -2.truept\hbox{\rlap{\hbox{$\sim$}}\raise5.truept \hbox{$<$}\ }}
\newcommand{\simg}{\raise
  -2.truept\hbox{\rlap{\hbox{$\sim$}}\raise5.truept \hbox{$>$}\ }}

\title[Hot Atmospheres in TNG100 and TNG50]{X-ray Signatures of Black Hole Feedback: Hot Galactic Atmospheres in IllustrisTNG and X-ray Observations}

\author[N. Truong et al]{Nhut Truong$^1$\thanks{truongnhut@caesar.elte.hu},
    Annalisa Pillepich$^{2}$,
    Norbert Werner$^{1,3,4}$,
    Dylan Nelson$^5$, \newauthor
    Kiran Lakhchaura$^{1,8}$,
    Rainer Weinberger$^6$,
    Volker Springel$^5$,
    Mark Vogelsberger$^7$,\newauthor
    and Lars Hernquist$^6$
 \\~\\
\footnotesize
$^1$ MTA-E\"otv\"os University Lend\"ulet Hot Universe Research Group, P\'azm\'any P\'eter s\'et\'any 1/A, Budapest, 1117, Hungary \\
$^2$ Max-Planck-Institut f\"ur Astronomie, K\"onigstuhl 17, D-69117 Heidelberg, Germany\\
$^3$Department of Theoretical Physics and Astrophysics, Faculty of Science, Masaryk University, Kotl\'a\v{r}sk\'a 2, Brno, 611 37, Czech Republic \\
$^4$School of Science, Hiroshima University, 1-3-1 Kagamiyama, Higashi-Hiroshima 739-8526, Japan \\
$^5$Max-Planck-Institut f{\"u}r Astrophysik, Karl-Schwarzschild-Str. 1, D-85748, Garching, Germany \\
$^6$Institute for Theory and Computation, Harvard-Smithsonian Center for Astrophysics, 60 Garden Street, Cambridge, MA 02138, USA \\
$^7$Department of Physics, Massachusetts Institute of Technology, Cambridge, MA 02139, USA \\
$^8$MTA-ELTE Astrophysics Research Group, P\'azm\'any P\'eter s\'et\'any 1/A, Budapest, 1117, Hungary
}

\begin{document}
\maketitle

\begin{abstract}
Hot gaseous atmospheres that permeate galaxies and extend far beyond their stellar distribution, where they are commonly referred to as the circumgalactic medium (CGM), imprint important information about feedback processes powered by the stellar populations of galaxies and their central supermassive black holes (SMBH). In this work we study the properties of this hot X-ray emitting medium using the IllustrisTNG cosmological simulations. We analyse their mock X-ray spectra, obtained from the diffuse and metal-enriched gas in TNG100 and TNG50, and compare the results with X-ray observations of nearby early-type galaxies. The simulations reproduce the observed X-ray luminosities ($L_{\rm X}$) and temperature ($T_{\rm X})$ at small ($<R_{\rm e}$) and intermediate ($<5R_{\rm e}$) radii reasonably well. We find that the X-ray properties of lower mass galaxies depend on their star formation rates. In particular, in the  magnitude range where the star-forming and quenched populations overlap, $M_{\rm K}\sim-24$~$ (M_*\sim10^{10.7}M_\odot)$, we find that the X-ray luminosities of star-forming galaxies are on average about an order of magnitude higher than those of their quenched counterparts. We show that this diversity in $L_{\rm X}$ is a direct manifestation of the quenching mechanism in the simulations, where the galaxies are quenched due to gas expulsion driven by SMBH kinetic feedback. The observed dichotomy in $L_{\rm X}$ is thus an important observable prediction for the SMBH feedback-based quenching mechanisms implemented in state-of-the-art cosmological simulations. While the current X-ray observations of star forming galaxies are broadly consistent with the predictions of the simulations, the observed samples are small and more decisive tests are expected from the sensitive all-sky X-ray survey with {\it eROSITA}. 
\end{abstract}
\section{Introduction}
It is evident from observations that most if not all elliptical galaxies and many disk galaxies host a supermassive black hole at their centre (SMBH, see \citealt{kormendy.ho.2013, graham.2016} for reviews). By extracting energy from gas accretion, those SMBHs are the power engines of active galactic nuclei (AGN). Theoretical studies of galaxy formation have shown that AGN feedback plays an essential role in shaping many properties of massive galaxies (e.g. \citealt{springel.etal.2005, booth.schaye.2009, choi.etal.2015, weinberger.etal.2017}). Furthermore, it is also widely considered the most plausible mechanism for star formation quenching in massive galaxies (see \citealt{man.belli.2019} and references therein).

From an observational point of view, one of the most promising avenues to study the effects of AGN feedback is galactic hot atmospheres (or coronae, see \citealt{tumlinson.etal.2017} and \citealt{werner.etal.2019} for reviews). The hot, diffuse, and soft X-ray emitting gas, that permeates the inter-stellar medium (ISM, within a few kpc from a galaxy center) or in many cases extends well beyond the stellar distribution to the circumgalactic medium (CGM, up to hundreds of kpc in galactocentric distance), encrypts important information about galaxy formation:  it reflects the complex interplay of various heating/cooling processes such as gravitational heating via virial shocks, radiative cooling, feedback from stellar activity  (e.g. supernovae explosions) and AGN. Unlike the hot intra-cluster medium, which extends to larger scales, the galactic hot atmospheres are closer to the sites of star formation and supermassive black hole activity and therefore their thermal properties are anticipated to be more sensitive to non-gravitational processes, including BH-driven feedback.
This is supported by recent X-ray observations of massive early-type galaxies (ETGs) with the {\it Chandra} X-ray observatory (e.g. \citealt{goulding.etal.2016,babyk.etal.2018,lakhchaura.etal.2018}), which show that the X-ray scaling relations, e.g. the X-ray luminosity-temperature ($L_{\rm X}-T_{\rm X}$) relation, are steeper than the self-similar predictions based only on gravitational heating. Those studies suggest that, while the hot gas properties of massive galaxies are primarily determined by the gravitational potential, they are also affected significantly by AGN feedback.  

The emerging consensus picture is that the hot atmospheres are stabilized by mechanical (also called radio mode) feedback driven by SMBHs at low accretion rates (e.g. \citealt{nulsen.etal.2009, randall.etal.2011, randall.etal.2015, hlavacek-larrondo.etal.2015}). For example, \cite{nulsen.etal.2009} studied a sample of 24 elliptical galaxies obtained from the {\it Chandra} archive and found that the jet power that is determined based on the X-ray cavities exceeds the luminosity of the cooling atmosphere (see Fig. 1 in their work). This result indicates that mechanical AGN feedback in the form of cavities may be sufficient to offset the energy lost due to the radiative cooling of the atmosphere.

Low-mass galaxies are generally expected to host fainter X-ray atmospheres, because their potential wells are shallower. However, observational studies of hot gas atmospheres in systems down to below the mass of the Milky Way (e.g. \citealt{strickland.etal.2004,tullmann.etal.2006, yamasaki.etal.2009,mineo.etal.2012, bogdan.etal.2013, bogdan.etal.2015,bogdan.etal.2017,li.wang.2013, li.etal.2017}) show that many spiral, late-type galaxies (LTGs) host detectable luminous X-ray atmospheres, with $L_{\rm X}\sim10^{40}$ erg~s$^{-1}$. In this low-mass regime, in addition to AGN feedback, the hot gas content is also expected to be influenced by stellar feedback (e.g. \citealt{crain.etal.2010,dave.etal.2011, dave.etal.2012, vandevoort.etal.2016,christensen.etal.2016, sokolowska.etal.2018}). Therefore, X-ray observations of these lower-mass systems potentially probe processes connected to their star formation status.

Cosmological simulations that include AGN feedback, e.g. \cite{McCarthy.etal.2010} (OWLS), \cite{lebrun.etal.2014} (cosmo-OWLS), \cite{planelles.etal.2014}, \cite{choi.etal.2015}, \cite{ liang.etal.2016}, \cite{henden.etal.2018} (FABLE), \cite{dave.etal.2019} (SIMBA), reproduce the hot gas properties in better agreement with observations than simulations that do not consider SMBH feedback. For example, \cite{choi.etal.2015} show that, without the inclusion of AGN feedback, simulations overestimate the X-ray luminosity of the hot atmospheres by more than 2 orders of magnitude compared to observations. More importantly, by comparing simulations with various treatments of AGN feedback (e.g. thermal versus mechanical), they point out that, in their implementation, the mechanical feedback is the responsible channel for reproducing the observed X-ray luminosities. However, their work is based on zoom-in simulations of a relatively small sample of 20 simulated galaxies, with exclusive focus on the high-mass end: $M_*>8.8\times10^{10}M_\odot$. 

In this paper, we aim to explore the hot galactic atmospheres using a large sample of simulated galaxies taken from the IllustrisTNG project (TNG; \citealt{nelson.etal.2018,naiman.etal.2018, marinacci.etal.2018,pillepich.etal.2018a,springel.etal.2018, pillepich.etal.2019, nelson.etal.2019}). In particular, in this work we use the TNG100 and TNG50 flagship runs (see Section~\ref{sec:sims} for a detailed description): these cover simulated volumes of $\sim(110\ {\rm Mpc})^3$ and $\sim(50\ {\rm Mpc})^3$, respectively, comparable to the volumes probed by current X-ray observations in the local Universe, and they have a numerical mass resolution good enough for us to confidently study systems down to the scale of $M_*\gtrsim$ a few $10^{9}M_\odot$. 

By construction, the TNG simulations are based on a galaxy formation model whose unconstrained choices have been adopted to reproduce observed stellar properties, e.g. the galaxy stellar mass function at $z=0$ (see \citealt{pillepich.etal.2018}). However, other outcomes, such as the temperature and metallicity of the hot gaseous atmospheres, are predictions of the simulation that can be readily compared with observations. For this task, we employ a dataset of $\sim160$ nearby galaxies that have {\it Chandra} and {\it XMM-Newton} X-ray observations in the literature (\citealt{mineo.etal.2012,li.wang.2013,li.etal.2017,goulding.etal.2016,babyk.etal.2018,lakhchaura.etal.2019}). In particular, in this paper we use the TNG simulations to get insights into the role that SMBH feedback can have 1) on shaping the X-ray properties of the gaseous atmospheres in galaxies across more than 2 orders of magnitude in stellar mass and 2) on the relationship between star formation quenching and gas content.

Within the TNG framework, earlier works by \cite{weinberger.etal.2017, weinberger.etal.2018,nelson.etal.2018, terrazas.etal.2019, davies.etal.2019} find a close connection between the suppression of star-formation rate in massive galaxies and the BH feedback in kinetic mode, whereby suggesting the crucial role played by the latter in establishing the quenched population in the TNG simulations. Here, we explore the connection between BH feedback, gas content and star formation activity by characterizing the hot atmospheres of star-forming and quenched galaxy populations in TNG100 and TNG50 at $z=0$, after having compared the X-ray properties of the hot atmospheres of simulated ETGs with observed ones. For this purpose, we perform mock {\it Chandra} X-ray observations of the simulated galaxies to mimic the typical observation procedure applied to the observed samples elected for the comparison. Importantly, throughout the paper, the X-ray signals we are interested in are produced by the diffuse, hot, metal-enriched gas in both simulations and observations: therefore, by construction, these signals do not account for the contribution from black holes, supernova remnants or binary stars that instead are a non-negligible contribution to the X-ray emission from the interstellar-medium of galaxies.

The paper is arranged as follows. We first describe in Section \ref{sec:2} the observed and simulated galaxy samples and the analysis of the mock X-ray observations. Section \ref{sec:3} is dedicated to the comparison between the simulated and observed ETG samples. We first describe the way we select analog quiescent galaxies from the simulations and perform a detailed comparison between the simulated and observed X-ray relations. Next, in Section \ref{sec:predictions}, we inspect the dependence of the X-ray luminosity on the galaxy properties, in particular galaxy star formation rate, for both simulations and observations. In Section \ref{sec:5}, we carry out a theoretical investigation on the origin of the difference in $L_{\rm X}$ in connection with the star formation rate and with the SMBH feedback. Finally, we conclude in Section \ref{sec:6}.   
\section{Methodology and galaxy samples}
\label{sec:2}

\begin{table*}
  \caption{\label{tb1}
  Summary of the observed datasets used in our study to contrast observational findings to the outcomes of the TNG100 and TNG50 simulations.}
 \begin{center}
  \resizebox{0.99\textwidth}{!}{
 \begin{tabular}{|cccccccc|}
 \hline
Dataset & Number of galaxies & Type & Distance & Aperture & Main Instrument & Exposure & Fitting Model \\
\hline
 MASSIVE (\citealt{goulding.etal.2016}) & 33 & Early-type & $<108$ Mpc & $R_{\rm e}$ & {\it Chandra} ACIS-S & 2-300 ks & APEC \\
 $\rm{ATLAS}^{3D}$ (\citealt{goulding.etal.2016}) & 41 & Early-type & $<42$ Mpc & $R_{\rm e}$ & {\it Chandra} ACIS-S & $>10$ ks & APEC \\
 \cite{lakhchaura.etal.2019} & 24 & Early-type & $<100$ Mpc & $R_{\rm e}$ & {\it Chandra} ACIS-S & 2-175 ks & APEC (version 3.0.7) \\
  \cite{babyk.etal.2018} & 42 & Early-type & $\siml150$ Mpc & $5R_{\rm e}$ & {\it Chandra} ACIS-S & $>10$ ks & APEC (version 3.0.7) \\
   \cite{mineo.etal.2012} & 20 & Late-type & $<40$ Mpc & $D_{25}$ & {\it Chandra} ACIS-S & $\geq$ 15 ks & MEKAL \\
  \cite{li.wang.2013} & 39 & (29) Late/(10) early-type & $\siml30$ Mpc & $D_{25}$ & {\it Chandra} ACIS-S & $\simg10$ ks & MEKAL/VMEKAL \\ 
  \cite{li.etal.2017} & 6 & Late-type & $<100$ Mpc & $30-100$ kpc & {\it XMM-Newton} EPIC & 45-123 ks & APEC \\
  \hline
 \end{tabular}}

 \end{center}
 \end{table*}
\subsection{The X-ray observational samples of reference}
\label{sec:obs}
In this paper, we compare the output of the TNG simulations (see Section~\ref{sec:sims}) to results from observations. In particular, we collect a number of galaxy datasets with available X-ray measurements. Their basic properties are summarized in Table \ref{tb1}, including e.g. subsets of the MASSIVE and $\ATLAS$ samples by \citealt{goulding.etal.2016}. 

To validate the TNG model in terms of gaseous atmospheres, we collect samples of massive early-type galaxies from observations. In particular, we employ the sample compiled by \cite{goulding.etal.2016}, which consists of 74 ETGs obtained from the MASSIVE and $\ATLAS$ surveys with available \textit{Chandra} X-ray observations. As the selection of these is based on well-defined optically-based criteria and their X-ray data are analyzed in a homogeneous way, we choose the \cite{goulding.etal.2016} sample as a reference throughout the paper. Below we briefly describe the selection as well as the X-ray analysis of the compiled MASSIVE and ${\rm ATLAS^{3D}}$ samples. 

Unlike in simulations, observations rarely come along with dynamical mass measurements, thus one has to rely on mass proxies. One of such proxies that is widely used in observational studies is the K-band absolute magnitude ($M_{\rm K}$) for it is considered to be closely linked to stellar mass (e.g. \citealt{cappellari.etal.2011,ma.etal.2014}). For all the observational datasets used in this paper, the K-band magnitude is collected from the Two Micron All Sky Survey (2MASS) database for extended sources (\citealt{skrutskie.etal.2006})\footnote{https://irsa.ipac.caltech.edu/applications/2MASS/PubGalPS/}. The total K-band magnitude is computed from the total K-band luminosity of the galaxy derived from a combination of the measured inner surface brightness profile and an extrapolated profile at larger radii obtained by fitting the inner one to a single Sersic profile (\citealt{jarrett.etal.2003}). 

The MASSIVE survey targets the most massive ETGs in the local Universe within a distance of $d\lesssim108$ Mpc and with an absolute K-band magnitude $M_{\rm K}<-25.3$ ($M_*\simg 10^{11.5}M_\odot$) resulting in a volume-limited sample of 118 galaxies (see \citealt{ma.etal.2014} for an overview). About $1/4$ of the original sample, i.e. 33 galaxies, have X-ray observations in the {\it Chandra} archive. The $\ATLAS$ survey, on the other hand, is dedicated to lower-mass ($M_{\rm K}<-21.5$ or $M_*\simg6\times10^9M_\odot$) galaxies, within a smaller distance of $d\lesssim42$ Mpc \citep[see][ for an overview]{cappellari.etal.2011}. 41 galaxies, out of the original sample of 260 nearby ETGs, have X-ray data obtained by {\it Chandra}. For both MASSIVE and $\ATLAS$, the early-type nature of the galaxies is established by selecting objects based on their morphology, i.e. only ellipticals and lenticulars (S0) are selected. In practice, this has been achieved by excluding galaxies with spiral arms upon visual inspection of their stellar-light images. X-ray spectra are extracted within a circular region of the half-light radius ($R_{\rm e}$) to allow direct measurements of X-ray quantities of the hot gas hosted by the galaxies. A model of a single-temperature plasma in collisional ionisation equilibrium (APEC, \citealt{smith.etal.2001}) is used to describe the X-ray emission of the hot inter-stellar medium. The spectral fitting for the temperature is limited to the energy range $[0.3-7]$ keV, while the X-ray luminosity is computed in the $[0.3-5]$ keV range.
 
Though starting with volume-limited and magnitude-selected samples, the X-ray subsets of MASSIVE and $\ATLAS$ are not complete. To enlarge the X-ray sample size, we therefore also consider 24 ETGs from \cite{lakhchaura.etal.2019} (out of an original sample of 47 nearby ETGs), with available {\it Chandra} observations and K-band measurements, which do not overlap with the MASSIVE+$\ATLAS$ sample. In addition, in order to constrain the X-ray relations across larger galactic apertures, we employ 87 ETGs obtained from \cite{babyk.etal.2018}, of which 45 systems overlap with the MASSIVE+$\ATLAS$+\cite{lakhchaura.etal.2019} sample. The X-ray properties of these galaxies were measured by {\it Chandra} within a radius of $5R_{\rm e}$. 

Finally, in order to investigate the X-ray properties of the hot gas in low-mass, star-forming galaxies and to compare our findings from TNG with existing observations, we also collect a sample of nearby late-type galaxies taken from \cite{mineo.etal.2012, li.wang.2013,li.etal.2017}. \cite{mineo.etal.2012} report an X-ray study of 20 late-type (spiral and irregular), star-forming galaxies at $d\lesssim40$~Mpc observed by {\it Chandra}, covering a broad range in star formation rates ($\sim0.1-17M_\odot\rm{yr}^{-1}$) and stellar masses ($\sim3\times10^{8}-6\times10^{10}M_\odot$). We also include a sample of 39 highly-inclined ($i\gtrsim60^{o}$), mostly late-type galaxies (see Table \ref{tb1}) at $d\lesssim30$~Mpc observed by {\it Chandra} and analysed by \cite{li.wang.2013}. The studied sample covers a range of about 2 orders of magnitude in stellar mass ($\sim10^{9}-10^{11}M_\odot$). Furthermore, we added 6 massive ($M_*\gtrsim10^{11}M_\odot$) spiral galaxies at $d\lesssim100$~Mpc observed by {\it XMM-Newton}  \citep{li.etal.2017}. For all these late-type galaxies, the systematic analysis of point source contamination, which is essential for X-ray studies of the hot gas of star-forming galaxies, is addressed extensively.
\subsection{The IllustrisTNG simulations}
\label{sec:sims}
The simulated galaxies used in this study are obtained from the IllustrisTNG\footnote{http://www.tng-project.org} project, a set of cosmological magneto-hydrodynamical simulations (\citealt{nelson.etal.2018,naiman.etal.2018, marinacci.etal.2018,pillepich.etal.2018a,springel.etal.2018}).
These are performed with the {\sc arepo} code (\citealt{springel.2010}), include a wide range of astrophysical processes, and are run with cosmological parameters consistent with results from Planck observations \citep{planck.2016}: matter density $\Omega_{\rm m}= 0.3089$, baryon density $\Omega_{\rm b}=0.0486$, dark energy density $\Omega_\Lambda=0.6911$, Hubble constant $H_0=67.74\ \rm{km\ s}^{-1}\rm{Mpc}^{-1}$, power spectrum normalization characterized by $\sigma_8=0.8159$, and primordial spectral index $n_{\rm s}=0.9667$.

The TNG model of galaxy formation \citep{weinberger.etal.2017, pillepich.etal.2018} is based on the original Illustris model (\citealt{vogelsberger.etal.2013,torrey.etal.2014}) and includes primordial and metal-line radiative cooling, prescriptions for star formation and evolution, supernovae feedback, metal enrichment, and supermassive black hole growth and feedback. 

The TNG suite includes runs with various volumes and resolutions. For this work, we use two simulated samples of galaxies at $z=0$ extracted from the TNG100 and TNG50 flagship runs. TNG100 covers a cosmological comoving volume of $(110.7\ {\rm Mpc})^3$ with a baryon mass resolution of $m_{\rm baryon}=1.4\times10^{6}M_\odot$. The recently-completed TNG50 (\citealt{pillepich.etal.2019, nelson.etal.2019}) offers the best resolution amongst the TNG simulations, with 16 times better mass resolution than TNG100 over a volume of $(51.7\ {\rm Mpc})^3$. Moreover, within the star-forming regions of TNG50 galaxies, the spatial resolution of the gas cells lies in the 70-140 pc range, with stellar and dark matter softening below 300 pc \citep[see Table 1 and Figure 1 of][ for more detail]{pillepich.etal.2019}. We combine the two datasets in a complimentary way, as TNG100 is optimal for probing the most massive galaxies, while TNG50 is necessary when studying low-mass galaxies. 


\subsubsection{SMBH growth and their feedback in the TNG model}  
Of particular importance for this study is the implementation of BH physics, which we hence summarize here \citep[see][ for more detail]{weinberger.etal.2017,weinberger.etal.2018}. 
For any Friend-of-Friend (FoF) halo identified on the fly with mass larger than $7.38\times 10^{10}M_\odot$ and with no BH yet, a SMBH with a mass of $1.18\times10^{6}M_\odot$ is seeded. Thus, the SMBH can grow by accretion of gas via an Eddington-limited Bondi model (see equations 1-3 in \citealt{weinberger.etal.2018}) or via merging with other SMBHs following the merging of their host galaxies.

For the modelling of BH feedback, the TNG model employs a two-mode scenario in which SMBHs can release feedback energy into the surrounding environment either in the form of thermal energy (thermal mode) or kinetic energy (kinetic mode). The total amount of the injected energy depends on the accretion rate onto the SMBH (see equations 7-9 as well as the corresponding numerical values for the efficiency parameters in \citealt{weinberger.etal.2017}). In the TNG model, the division into the two modes is controlled by the BH accretion rate. While the thermal mode is present when SMBHs are at high accretion rates, the kinetic mode is switched on when the value of the Eddingtion ratio drops below the threshold:
\begin{equation}
    \chi={\rm min}\bigg[0.002\bigg(\frac{M_{\rm BH}}{10^8M_\odot}\bigg),0.1\bigg], \label{eqn1}
\end{equation}
where $M_{\rm BH}$ is the SMBH mass. The numerical values in equation (\ref{eqn1}) are determined so that the TNG simulated galaxies show realistic properties in terms of their stellar component, such as the stellar mass function and spatial extent of the stellar population at $z=0$.

\subsection{Measurement of observables from simulated galaxies}
 In this Section we describe the definitions as well as the procedures used to compute observable quantities from simulated data. 
 \subsubsection{Mock X-ray analysis and Intrinsic X-ray properties}
\label{sec:mocks}
 For comparison with X-ray observations, we carry out mock X-ray analyses of simulated galaxies to obtain values for the X-ray luminosities and the gas temperatures that closely mimic those determined from actual observations. Here we are after the X-ray signal produced by the diffuse, hot, metal-enriched gas within and around galaxies and we deliberately neglect the X-ray contribution from point-like sources such as black holes, supernova remnants or X-ray binaries -- which in fact are not modeled explicitly in the simulations.
 
 In practice, the analysis involves two steps: i) generating mock spectra for a collection of gas cells within a region of interest for each simulated galaxy, and ii) fitting the integrated mock spectra to obtain X-ray quantities (such as the X-ray gas temperature, $T_{\rm X}$, and the X-ray luminosity, $L_{\rm X}$) on a galaxy by galaxy basis.  
 
For any given object, the gas cells are selected from a cylindrical region that is randomly oriented with respect to the galaxy structure -- in our case, along the z-axis of the simulation box -- and is centered at the galaxy position -- i.e. the location of its most gravitationally bound resolution element as determined by the {\sc subfind} halo finder. To account for projection effects, the cylinder height is equal to $10\times R_{\rm e}$, where $R_{\rm e}$ is the effective radius (or half light radius, see the definition below) and we measure the X-ray signals within projected circles with $1$ or $5\times R_{\rm e}$ radii. Only non star-forming gas cells are used, though we have verified that including star-forming gas cells increases the total X-ray emission by an insignificant amount ($\lesssim 1/1000$ per galaxy), because the X-ray emission of low-temperature gas cells ($T<10^{10.5}$ K) is negligible. We discuss the contribution from the subgrid hot components of the star-forming gas cells in Appendix \ref{sec:appC}. Moreover, only gas cells that are gravitationally bound to the galaxy of interest, which are identified by the {\sc subfind} algorithm, are considered. In the case of central galaxies, our mock X-ray signals therefore automatically excise the contribution from e.g. satellite galaxies, as typically done in observations. In the case of satellite galaxies orbiting in more massive groups and clusters, our mock X-ray measurements naturally exclude the contribution from the background ICM. However, in the case of a central galaxy, in our mocks no additional contribution is subtracted off, namely we do not model and exclude a possibly separate contribution from the ICM, as sometimes done in observations. 

Note that within this framework, the X-ray contribution from outflowing gas may not be fully accounted for. On the one hand, by construction, wind particles, i.e. those that are launched within the TNG non-local stellar feedback scheme (see \citealt{pillepich.etal.2018} for detailed description of the stellar feedback model), cannot contribute at all to the X-ray signal measured here, but it should be noted that wind particles are spawned from star-forming regions within galaxies, so from cold gas that should not contribute in any case to the X-ray signal. On the other hand, high-velocity gas outflows may not be accounted for within the adopted setup because they may be missed by the {\sc subfind} algorithm as they may not be gravitationally bound to their host halo due to their fast moving nature: such cells may span a wide range of temperatures and densities and may in principle contribute to the X-ray signals. We have attempted to examine the relative contribution of high-velocity gas components by inspecting the comparison between X-ray emissions that come from {\sc subfind}-selected gas cells and those identified by the Friend-of-Friend algorithm. This test is suitable for central galaxies only: we find that at the high-mass end ($M_{\rm K}<-24$) there is no significant difference between X-ray luminosities (within $R_{\rm e}$) that are measured based on {\sc subfind}- and FoF-selected gas cells, while at the low-mass end the difference is more noticeable, in particular for star-forming galaxies whereby the FoF-selected X-ray emission is on average a factor of $\sim2$ higher than the {\sc subfind}-selected value. Nonetheless, we notice that the difference is significantly smaller than the intrinsic scatter of the X-ray luminosity of central galaxies at the same mass range\footnote{The $84^{th}$ percentile value of the X-ray luminosity ($<R_{\rm e}$) for central galaxies with $M_{\rm K}>-24$ is a factor of $\sim12$ higher than the median value.}. On a different note, for satellite galaxies, this test shows that the FoF-cell-based measurements would be heavily contaminated by the signal from the gas of the central galaxy of the same host halo, thereby supporting our choice to adopt the gravitationally-bound material as the source of the X-ray signals for both centrals and satellites.
\begin{figure*}
    \centering
    \includegraphics[width=0.99\textwidth]{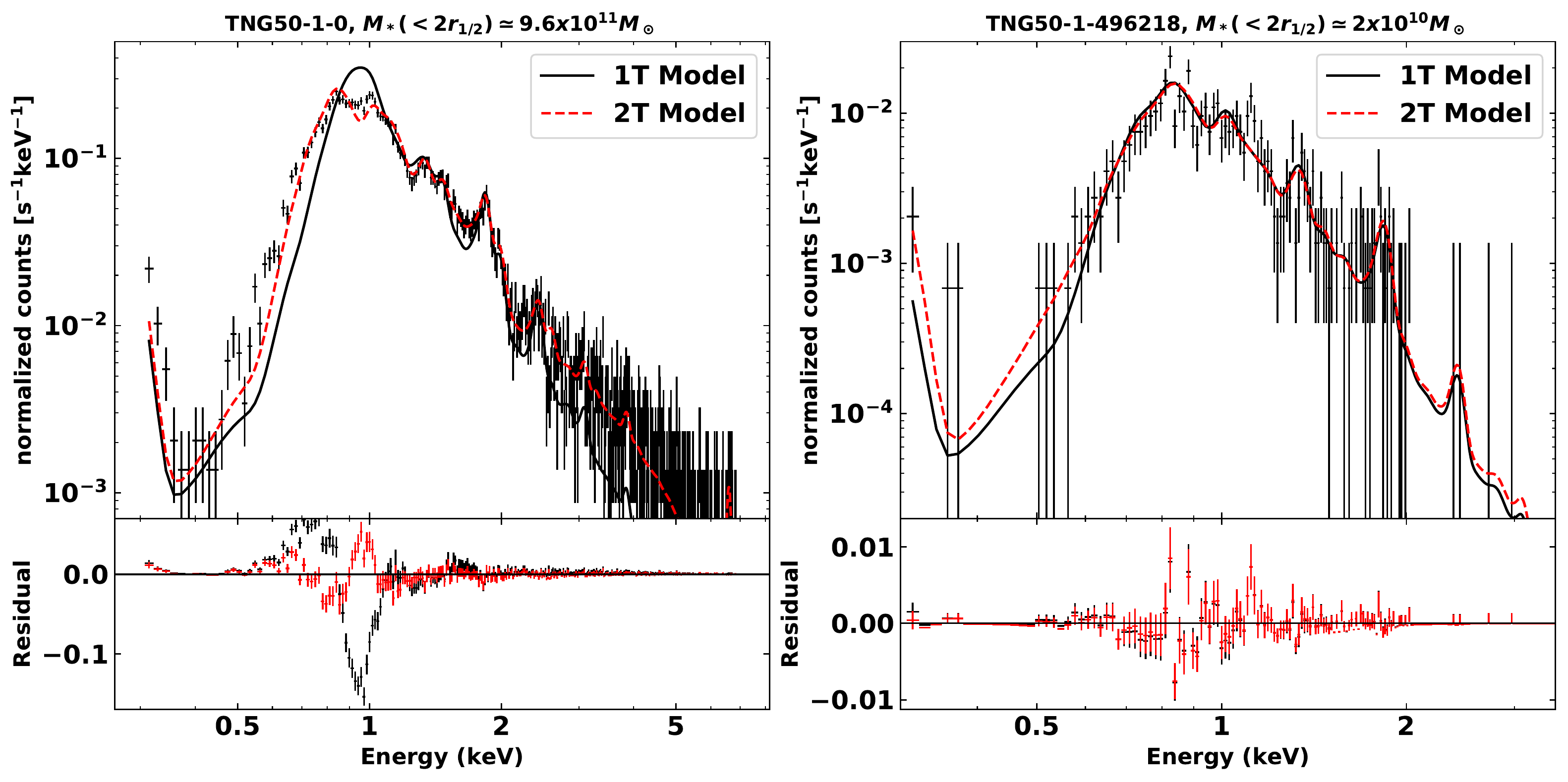}
\caption{Examples of mock X-ray spectra and best-fitting curves with one- (1T) and two-temperature (2T) APEC models for a high-mass (left) and a low-mass (right) galaxy taken from the TNG50 simulation at $z=0$ (see text in Section \ref{sec:mocks} for a detailed description of the mock X-ray analysis).}
    \label{fig:1}
\end{figure*}

For each gas cell, a mock spectrum is generated based on its gas density, temperature, and metallicity assuming a single temperature APEC model (version 3.0.9) plus galactic absorption (``wabs(apec)'') using the \texttt{fakeit} procedure that is implemented in the  XSPEC\footnote{https://heasarc.gsfc.nasa.gov/xanadu/xspec/} (\citealt{smith.etal.2001}) package. The mock spectrum is created for an exposure time of $100$ ks, which is consistent with the typical depth of actual observations (see Tab. \ref{tb1}), no background is applied and statistical errors are assumed to be Poisson, based on the photon counts. We use a column density $n_{\rm{H}}=10^{20}\rm{cm}^{-2}$.  To be more realistic, the simulated spectra are convolved with the response files of {\it Chandra}\footnote{We use response files for the default pointing of the 20th-cycle {\it Chandra} ACIS-S detector.}, assuming an energy resolution of $150$ eV. The final spectrum is obtained by adding up all spectra created from the gas cells belonging to the given region of interest. The mock spectra are then fitted assuming either a single temperature APEC model (1T, ``wabs(apec)'') or a two-temperature model (2T, ``wabs(apec+apec)''), by using the counts in the $[0.3-7]$ keV range and by fixing the galaxy metallicity to its emission-weighted values\footnote{$Z_{\rm ew}=\frac{\sum_{\rm i}\epsilon_{\rm i}\times Z_{\rm i}}{\sum_{\rm i}\epsilon_{\rm i}}$, where $Z_{\rm i}$ is the metallicity of the $i^{th}$ gas cell and $\epsilon_{\rm i}$ is the X-ray emission computed as described in equation (\ref{eqn2}).} (see Appendix \ref{sec:appB} for a detailed inspection). We adopt the solar abundances values provided by \cite{anders.grevesse.1989}. The fit thus returns best-fitting values and associated uncertainties for all parameters of the fitting model, namely the gas temperature(s) and the normalisation (which is proportional to the gas density squared). The X-ray luminosity is derived from the best-fit plasma model (APEC) for each simulated galaxy. 

For illustration of the mock X-ray analysis procedure, we show a couple of examples of mock X-ray spectra as well as their fits in Fig. \ref{fig:1} for two present-day galaxies at the high and the low-mass end from the TNG50 simulations. In general, a 1T model is sufficient to fit the X-ray spectra of hot gas in galaxies across the considered mass range, except in high-mass systems where a 2T model is often required to improve the fitting. Those represent the most massive galaxies at the center of groups or clusters with non-negligible temperature gradients that make a 1T model inadequate for the fit. Nonetheless, we have verified that, on average, the X-ray relations obtained with 1T or 2T models vary by an insignificant amount compared to their own intrinsic scatters. Therefore, for the rest of this paper, we opt to show results obtained from fitting the mock spectra with 1T models only.

To examine how reliable the mock X-ray analysis is, we compare its results with theoretical quantities that can be directly measured based on each gas cell properties, such as the gas temperature, density, and metallicity. 

The {\it intrinsic} total X-ray luminosity of a simulated galaxy is obtained by summing the X-ray emission from all the gas cells within a region of interest:
\begin{equation}
    L_{\rm X, intrinsic}=\sum_{\rm i} \epsilon_{\rm i}, \label{eqn2}
\end{equation}
where $\epsilon_{\rm i}$ is the X-ray gas emission in the $[0.3-5]$ keV band computed for the $i^{\rm th}$ gas cell assuming a 1T APEC model. 

Averaged gas temperatures can be measured using two different weights, the gas mass-weighted ($T_{\rm mw}$) and the emission-weighted ($T_{\rm ew}$), according to the formula:
\begin{equation}
    T_{\rm mw, ew}=\frac{\sum_{\rm i} w_{\rm i}\times T_{\rm i}}{\sum_{\rm i} w_{\rm i}}, \label{eqn3}
\end{equation}
where $T_{\rm i}$ is the $i^{th}$ gas cell temperature, and $w_{\rm i}=m_{\rm gas,i}$ (gas mass) for the case of $T_{\rm mw}$ and $w_{\rm i}=\epsilon_i$ for the case of $T_{\rm ew}$. 

As we explicitly show in Appendix~\ref{sec:appA}, it is not possible to obtain mock X-ray measurements for all simulated galaxies in our mass-limited samples. Especially at the low-mass end, galaxies may produce zero or a very low number of photons that are received by {\it Chandra} in a 100~ks exposure time. This may occur because of the limited numerical resolution or for actual physical reasons, e.g. the gas of the considered galaxies may be too cold to emit photons in the energy range of interest. In the case of a very low number of photons, it is not possible to obtain an X-ray temperature from spectral fitting. To avoid galaxies that do not produce a sufficient number of photons that in turn may result in bad or impossible fits and unreliable mock X-ray measurements, we flag systems with $T_{\rm X}$ lying beyond $3\sigma$ off the average $T_{\rm X}-T_{\rm ew}$ relation: these are excluded from the analysis. All results throughout the paper will therefore include only TNG galaxies with reliable mock X-ray measurements for both $T_{\rm X}$ and $L_{\rm X}$: these are labeled ``X-ray detected''. Approximately, the ``X-ray detected'' sample consists of galaxies that have $L_{\rm X}(<R_{\rm e})\gtrsim5\times10^{37}$ erg~s$^{-1}$. A more detailed discussion on this selection and about the comparison between mock X-ray quantities and theoretically computed quantities is given in Appendix~\ref{sec:appA}.
\begin{figure*}
    \centering
    \includegraphics[width=0.99\textwidth]{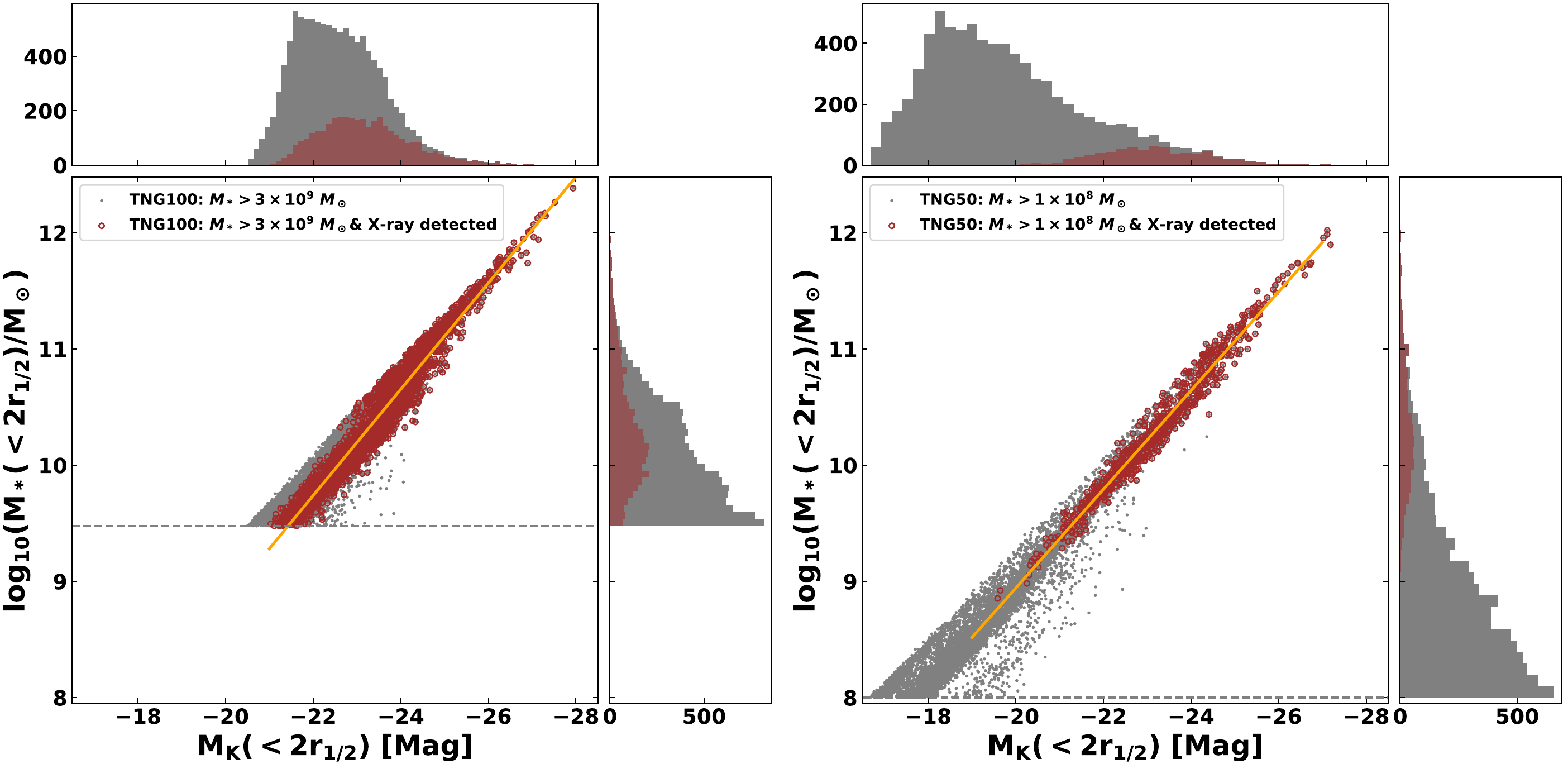}
\caption{The stellar mass-magnitude relation for TNG100 ({\it left}) and TNG50 ({\it right}) galaxies at $z=0$. The main panels show the scatter plots of stellar mass and magnitude, while the sub-panels represent the histograms of the two quantities. The grey data points denote simulated galaxies selected to have at least a minimum stellar mass (denoted by dashed lines): to ensure sufficient resolution of stellar properties. In red, we show galaxies for which we have reliable X-ray measurements. All plots in this paper are based on TNG100 and TNG50 denoted here by red points: they are objects that pass the mock X-ray analysis as presented in Section \ref{sec:mocks} and Appendix~\ref{sec:appA}. The solid lines represent the best-fitting $M_*-M_{\rm K}$ relations as provided in equations (\ref{eqn4}) and (\ref{eqn5}).} 
    \label{fig:2}
\end{figure*}

\subsubsection{Other galaxy properties}
\label{sec:props}
The hot gas properties of galaxies are expected to be inextricably causally related to their stellar and black hole activities. Therefore, beside computing X-ray quantities as described above, we also utilize other measurements to characterize TNG galaxies, specifically in relation to their stellar content and SMBH properties.
\begin{itemize}
    \item {\it Half-mass ($r_{1/2}$) and half-light ($R_{\rm e}$) radii.} The former refers to physical radius within which half of the total stellar mass of the galaxy is contained. The half-light radius, also called effective radius, is defined as the radius within which half of the stellar light of the galaxy is contained. In both cases, all gravitationally-bound stellar particles are considered for the size measurements, instead of e.g. accounting only for the light down to an effective surface brightness limit. In this work, in order to characterise the extent of hot atmospheres,  we use the 2D circularized projected half-light radii computed in the K-band (\citealt{genel.etal.2017}): these do not account for the effects of dust. 
    \item {\it Stellar mass} ($M_*$) is the mass in the stellar component measured within twice the half-mass radius (i.e. $<2r_{1/2}$). We use this for mere reference, and not for comparisons to observations.
    \item {\it K-band absolute magnitude} ($M_{\rm K}$) is computed from the total luminosity in the K-band of the stellar particles that lie within $2\times r_{1/2}$. It is noted that no dust attenuation is modelled, thus this quantity is aimed to be compared with the extinction-corrected $M_{\rm K}$ from observational data. In fact, as noted in Section~\ref{sec:obs}, observationally-derived values are obtained from integrating a galaxy's magnitude via extrapolation of the light with a single Sersic profile: we comment in the next Sections to what levels the mismatch of operational definitions impacts our simulation-observation comparison. 
    \item {\it Galaxy color ($u-r$)} is obtained from integrated stellar light measured within 30 kpc in SDSS-u and SDSS-r bands and accounting for the effects of dust (see \citealt{nelson.etal.2018} for detailed discussion). We use $u-r$ colors to separate the simulated samples into two classes:
        \begin{itemize}
        \item blue: $u-r \leq 2.1$;
        \item red: $u-r > 2.1$.
        \end{itemize}
    \item {\it Stellar morphology} refers to parameters that describe the 3D shape of the stellar distribution. Following \cite{pillepich.etal.2019} (see also \citealt{chua.etal.2019}), we use axis ratios (see \citealt{pillepich.etal.2019} for detailed description), to characterize stellar distribution, e.g. disky versus spheroidal or elongated galaxies.   
    \item {\it Specific star formation rate ($\rm{sSFR}$)} is defined as the ratio of the instantaneous star-formation rate to stellar mass, both within twice the stellar half mass radius: $\rm{SFR}(<2r_{1/2})/M_{*}(<2r_{1/2})$.
    \item {\it Star formation activity flags} are used to specify the star formation status of a galaxy. We employ the operational definitions taken from \cite{pillepich.etal.2019} to classify simulated galaxies based on their instantaneous star-formation rate (SFR) and the logarithmic distance with respect to the star-forming main sequence at the corresponding stellar mass ($\Delta\log_{10}({\rm SFR})$). More specifically, the following flags are used in this work:
    \begin{itemize}
        \item star-forming: $\Delta\log_{10}({\rm SFR})>-0.5$. 
        \item quenched: $\Delta\log_{10}({\rm SFR})\leq-1.0$.
    \end{itemize}
    \item {\it BH feedback-to-binding energy ratio} is defined as $E_{\rm kin}/E_{\rm bin}\equiv\int \dot{E}_{\rm kin}dt/E_{\rm bin}$, where the numerator is the accumulated kinetic feedback released by the central super-massive black hole, and the denominator is the gravitational potential energy computed for all the gas cells within $2\times r_{1/2}$ (see also \citealt{terrazas.etal.2019} for the use of this quantity).
\end{itemize}

\begin{figure*}
    \centering
    \includegraphics[width=0.99\textwidth]{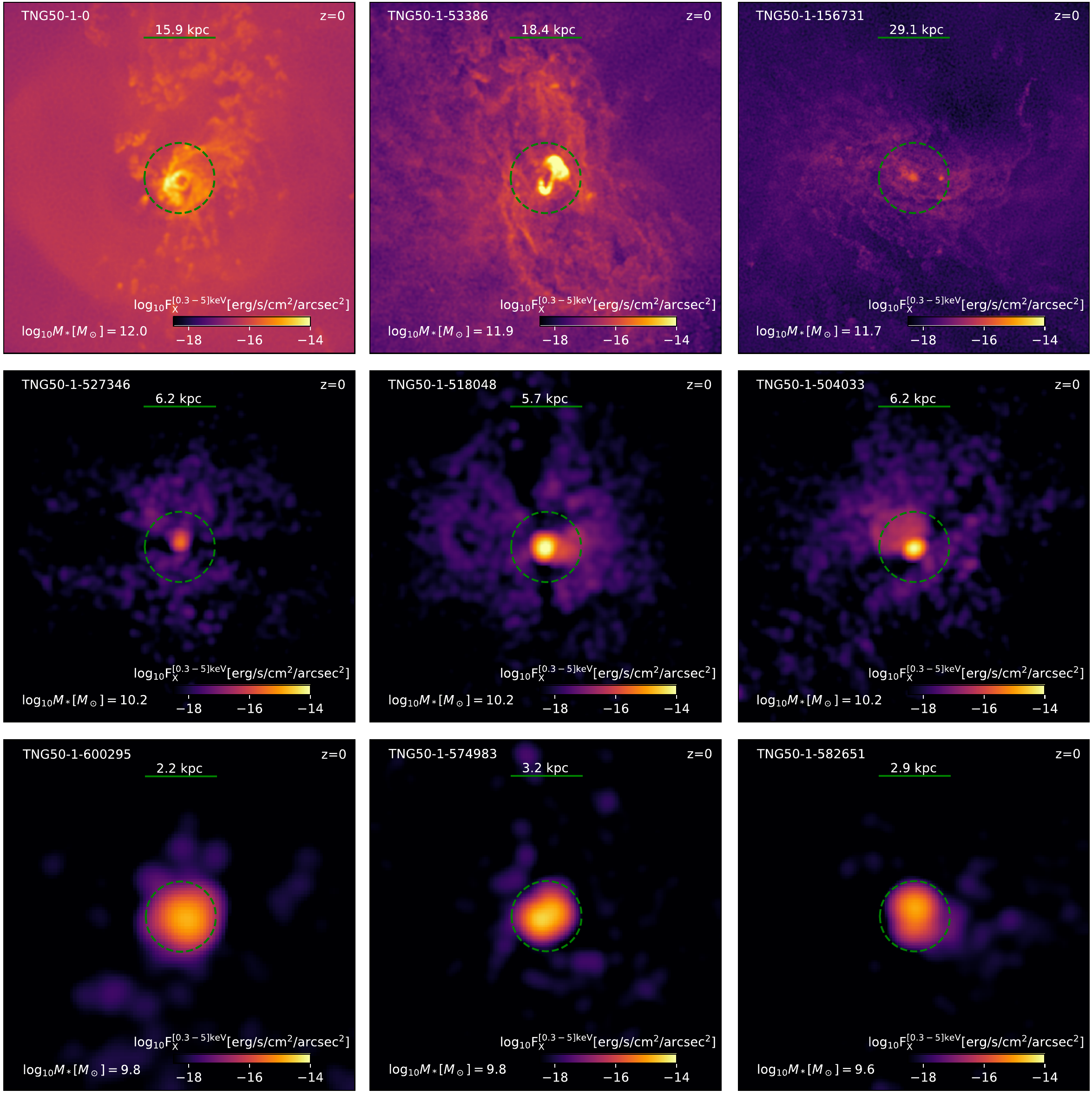}
\caption{X-ray flux maps for a random selection of TNG50 galaxies from $\sim10^{12}M_\odot$ ({\it top left}) to $\sim10^{9.6}M_\odot$ ({\it bottom right}) in stellar mass, at $z=0$. Each map has a size of $10R_{\rm e}\times10R_{\rm e}$ and the signal is integrated through a depth of $10R_{\rm e}$. Dashed circles represent regions within $R_{\rm e}$, the K-band half-light radius. The maps are created with assumed angular-diameter distance $D_{\rm A}\approx43.7$ Mpc ($z\sim0.01$).}
    \label{fig:3}
\end{figure*}
\subsection{The TNG100 and TNG50 simulated galaxies and their hot atmospheres}
\label{sec:tng}
Starting from {\sc subfind} haloes \citep{springel.etal.2001}, we select mass-limited samples of $z=0$ galaxies from TNG100 ($M_*>3\times10^9M_\odot$) and TNG50 ($M_*>10^8M_\odot$), so that in both simulations all galaxies are resolved with at least a few thousand stellar particles. Note that TNG50 has about 16 times better mass resolution than TNG100, albeit sampling a smaller volume: in the following, by simultaneously studying the hot atmospheres of both TNG100 and TNG50, we provide an estimate of how numerical resolution affects our quantitative results.

Throughout the paper, we consider both central and satellite galaxies, with no distinction. We verify that including satellites does not change significantly any of our qualitative or quantitative conclusions. Moreover, unless otherwise explicitly stated, we consider all galaxy types, independently of morphology, color or star formation state. 

These mass-selected samples account for $11233$ and $7540$ galaxies for TNG100 and TNG50, respectively, and are shown as gray dots in Fig. \ref{fig:2} on the $M_*-M_{\rm K}$ diagram. However, as mentioned in Section~\ref{sec:mocks} and described in Appendix~\ref{sec:appA}, not all galaxies produce enough photons in the $[0.3-7.0]~\rm{keV}$ energy band in a 100~ks exposure with {\it Chandra} to ensure a reliable spectral fit. In Fig. \ref{fig:2}, red data points indicate TNG galaxies with available and reliable mock X-ray properties (labeled as ``X-ray detected''): in practice, we find that simulated galaxies with a few $10^9\ M_\odot$ and above ($M_{\rm K}\siml-21$) start to host X-ray emitting gas (with $L_{\rm X}\gtrsim 5\times10^{37}$ erg~s$^{-1}$). The final sample of TNG100 (TNG50) X-ray-detected galaxies consists of 3523 (736) objects, which is about 31 per cent (10 per cent) of the original mass-selected sample\footnote{The fractional different between TNG100 and TNG50 is due to the different adopted minimum stellar mass cut: if we restricted the TNG50 sample to galaxies more massive than $M_*>3\times10^9M_\odot$, as for TNG100, more than 40 per cent would have well-defined X-ray measurements.}. Of the X-ray detected samples, about $67\%$ ($73\%$) are quenched galaxies, and about $21\%$ ($15\%$) are star-forming galaxies for TNG100 (TNG50), based on the star formation activity flags defined in Section~\ref{sec:props}. 

To validate the use of the K-band magnitude ($M_{\rm K}$) as a mass proxy, we examine the $M_*-M_{\rm K}$ relation for our sample of simulated X-ray bright galaxies and verify that their K-band magnitude does indeed correlate strongly with the stellar mass following the relations:
\begin{equation}
    \log_{10}M_*=10.2-0.47\times(M_{\rm K}+23),\ \ \rm{(TNG100)} \label{eqn4}
\end{equation}
\begin{equation}
    \log_{10}M_*=10.2-0.43\times(M _{\rm K}+23),\ \ \rm{(TNG50)} \label{eqn5}
\end{equation}
with an intrinsic scatter of $\sim0.1$ dex. We note that, while the slopes are consistent, the normalisation of the TNG relations is about $0.3$ dex smaller than the corresponding relation used for the MASSIVE sample (see equation 2 in \citealt{ma.etal.2014}). A possible reason for the offset lies in the different simulated and observed measurements of the stellar mass. The observed estimation is based on the work by \cite{cappellari.etal.2013} who approximated the stellar mass as $M_*\approx2M_{1/2}$ (see equation 28 in their paper), where $M_{1/2}$ is the total mass measured within the half-light radius from dynamical modeling. We verify in the TNG simulations that the same approximation would result in overestimating the stellar mass by $\sim0.3-0.6$ dex at $M_{\rm K}\sim -23$. This result further explains why we prefer to use a mass proxy, e.g. $M_{\rm K}$, for the comparison with observations, instead of stellar mass directly.

Finally, in Fig.~\ref{fig:3}, we show the X-ray maps of a selection of  TNG50 simulated hot atmospheres, from high (top left) to low masses (bottom right). We notice a marked diversity in the X-ray morphology across the sample. At the high-mass end ($M_*\gtrsim5\times10^{11}M_\odot$), the hot atmospheres appear relatively smooth, volume-filling, and they extend far beyond the stellar distribution ($\gtrsim 5R_{\rm e}$). This result is expected, as massive galaxies reside in massive haloes that maintain a stable accretion shock at the virial radius, which heats the accreted gas to the virial temperature (e.g. \citealt{birnboim.dekel.2003}). In addition to gravitational heating, previous studies of the TNG simulations show that in massive systems feedback powered by gas accretion onto the central SMBH is the dominant extra heating channel (\citealt{weinberger.etal.2018}), which can disperse the gas from the central regions \citep{terrazas.etal.2019}, heat it up (Zinger et al. in prep), and drive high-speed galactic outflows (\citealt{nelson.etal.2019}).

Moving toward lower masses, the hot atmospheres become less extended and less volume-filling, with the X-ray emission appearing more concentrated in the central regions ($\siml R_{\rm e}$).
In the TNG simulations, for galaxies with stellar mass below $10^{10}~M_\odot$, beside gravitational heating, stellar feedback is the main channel of extra energy (\citealt{weinberger.etal.2018}). In the middle row, for example, the hot atmospheres exhibit bipolar features, with the X-ray emitting gas extending beyond the galactic disk: these are indeed star-forming, disky galaxies and the cold, gaseous, star-forming disks appear as black ``edge-on'' regions.

\section{Comparison between TNG and Observed Early-Type Galaxies}
\label{sec:3}
ETGs have been the main focus of past X-ray observations because they have been considered on average massive enough to host hot atmospheres that emit abundantly in the X-ray band. Previous theoretical studies (e.g. \citealt{McCarthy.etal.2010,lebrun.etal.2014,planelles.etal.2014,choi.etal.2015,liang.etal.2016,henden.etal.2018,dave.etal.2019}) showed that the hot gas content in those massive galaxies is particularly susceptible to SMBH feedback, thereby making their X-ray observations an ideal avenue to constrain models of SMBH feedback. 
However, in order to make meaningful and quantitative comparisons, it is critical to define a sample of simulated galaxies that properly represents the observed sample of ETGs elected for the comparison. 
In this Section, we first describe the selection of a sample of ETG-like galaxies from TNG based on various optical properties and then compare X-ray relations, i.e. $L_{\rm X}-M_{\rm K}$ and $T_{\rm X}-M_{\rm K}$, between the selected simulated and observed datasets.

\begin{figure*}
    \centering
    \includegraphics[width=0.99\textwidth]{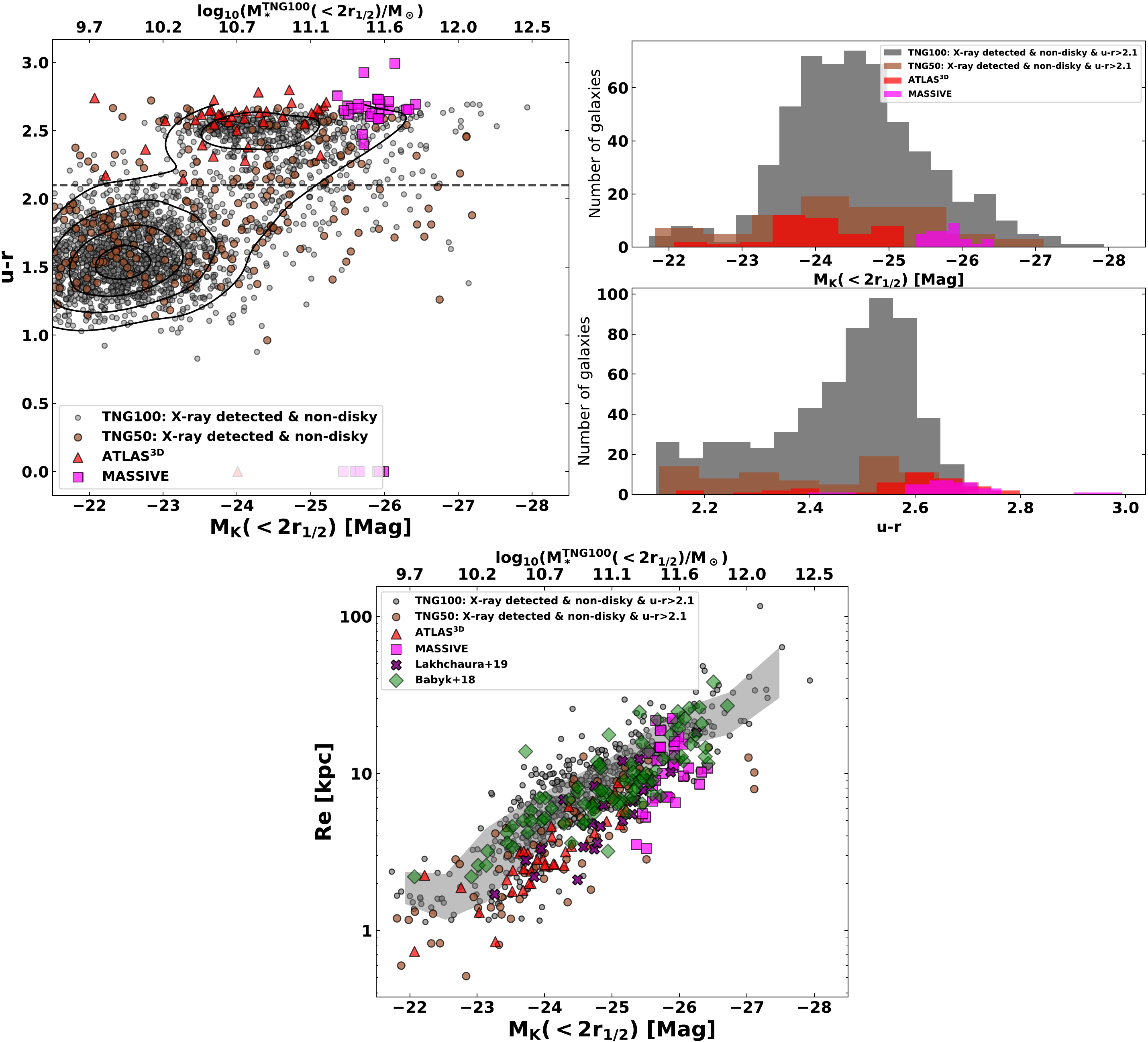}
\caption{Stellar-light properties of the simulated and observed samples of early-type galaxies at $z\simeq0$. In these panels, we only show those TNG100 (TNG50) galaxies with $M_*>3\times10^9M_\odot$ ($>10^8M_\odot$) that provide enough X-ray photons for our mock {\it Chandra} observations to measure reliable temperatures (see Section~\ref{sec:mocks} for detail). {\it Upper left:} the color-magnitude diagram for non-disky galaxies in TNG100 and TNG50 (grey and brown circles, respectively) is shown along with the corresponding observed samples of $\ATLAS$ and MASSIVE galaxies. The contours, computed for the TNG100 data, specify concentration levels of $90$, $60$, $30$, and $10$ per cent  of the galaxy distribution density maxima. Those observed galaxies that do not have a $u-r$ measurement are assigned a zero value of $u-r$. The dashed-line specifies the threshold of $u-r=2.1$ above which TNG galaxies are selected for the comparison with these observational datasets. {\it Upper right:} the two histograms of magnitude ($M_{\rm K}$, top) and color ($u-r$, bottom) are shown for the selected simulated sample (grey and brown histograms) as well as the observed sample. {\it Lower row:} comparison on the $R_{\rm e}-M_{\rm K}$ plane for the TNG galaxies selected as ``ETG-like'' and the observed datasets. The shaded area represents $1\sigma$ envelope about the median relation for the TNG100 sample.}
    \label{fig:4}
\end{figure*}
\subsection{Matching the simulated to the observed samples of ETGs}
\label{sec:matching}

Properly accounting for, and thus reproducing in the simulations, the selection of observed datasets is challenging: for example, the MASSIVE and $\ATLAS$ early-type galaxies are selected based on morphological criteria, i.e. they are ellipticals and S0s. Yet, such selection is not easily reproducible, because it is based on visual inspection, and more quantitative criteria may not return the same galaxy sample.

We attempt to select TNG galaxies that resemble the reference $\ATLAS$ and MASSIVE samples of ETGs from \cite{goulding.etal.2016} by imposing the following criteria:
\begin{itemize}
    \item K-band absolute magnitude ($M_{\rm K}$): we apply the magnitude cut $M_{\rm K}<-21.5$ to match the $\ATLAS$ low-mass threshold.
    \item Morphology: we select only non-disky simulated galaxies to mimic the observed sample of ellipticals and S0s. For this task, we apply the same criteria for the stellar axis ratios used in \cite{pillepich.etal.2019}, namely we consider as non-disky all those galaxies that do {\it not} satisfy the following properties: $q>0.66$ and $s<0.33$.
    \item Stellar color: we take red TNG galaxies, i.e. with $u-r>2.1$. 
\end{itemize}
These add to the requirements of having a minimum galaxy stellar mass and of being X-ray luminous (see Sections~\ref{sec:mocks} and \ref{sec:tng}).

The demographics of the simulated ETG-like galaxies are presented in the upper row of Fig.~\ref{fig:4} (grey and brown symbols for TNG100 and TNG50, respectively) in comparison to the compiled early-type sample of \cite{goulding.etal.2016}. There we show the color-magnitude (left) and the magnitude/color histograms (right). 

It is apparent from the color-magnitude diagram that the magnitude and morphological criteria alone are not adequate to disentangle quenched from star-forming galaxies in simulations. Both TNG100 and TNG50 non-disky galaxies display bimodal distribution in the $(u-r)-M_{\rm K}$ diagram. Therefore we opt to apply a color cut for the simulated objects, $u-r>2.1$, which matches the minimum value of the observed sample\footnote{The color data of the observed sample is collected from the NASA-Sloan Atlas database: www.nsatlas.org (see e.g. \citealt{blanton.moustakas.2009}).}. 

In the right panels, we inspect the magnitude and color distributions. Grey and brown histograms represent TNG100 and TNG50 ETG-like galaxies, i.e. non-disky and red objects, respectively. A couple of points are worth emphasising when it comes to discussing the comparison of hot gas properties later.

i) The selected TNG100 and TNG50 samples are more or less similar to the observed ones regarding the magnitude distribution except at the bright end and that the latter are somewhat overall flatter. In particular, in TNG100, the brightest simulated systems appear to be $\sim1.5$ magnitude brighter than the observed one. It is important to emphasize that it is difficult to replicate the exact $M_{\rm K}$ measurement performed for the observed data: in fact, we do not do that here, as the simulation magnitudes account for the stellar light from within twice the stellar half mass radius while the observed ones are obtained from extrapolating a single Sersic profile. From the observation side, there have been concerns (e.g. \citealt{lauer.etal.2007,ma.etal.2014}) that the relatively shallow photometry (the $1\sigma$ surface brightness limit is $20\ {\rm mag\ arcsec^{-2}}$) provided by the 2MASS survey might bias low the measurement of the K-band magnitude. As the radial range used for fitting the light profile is too small to obtain an accurate Sersic index, the total stellar luminosity could be underestimated especially for the cases of massive extended galaxies. On the other hand, from the simulation side, we have already compared TNG100 galaxy stellar mass functions at $z=0$ to observational results \citep{pillepich.etal.2018a}. We verify that when limiting the radius within which the simulated $M_{\rm K}$ is computed to a smaller value, e.g. $r=30$ kpc instead of $2\times r_{1/2}$\footnote{For $M_{\rm K}<-26$, the typical value of half-mass radius is $r_{1/2}\sim30$ kpc (see also the bottom plot of Fig. \ref{fig:4} for the values of the half-light radius, i.e. $R_{\rm e}$).}, the discrepancy in $M_{\rm K}$ between the brightest galaxies in TNG100 and observations is reduced to $\sim0.8$ magnitude. 

ii) For the colors, while the simulated distributions in the red region centre around the value of $u-r=2.5$, the observed sample is slightly shifted to a higher value of $\sim2.6$. Moreover, there are a couple of MASSIVE systems that are significantly redder than the simulated ones ($u-r\gtrsim3$). Finally, the simulated non-disky galaxies can extend to much lower values of the $u-r$ distribution. \citealt{nelson.etal.2018} have demonstrated that, across morphological types, the TNG100 galaxy population is in striking quantitative agreement with the SDSS $g-r$ and $u-r$ vs. mass distributions, but indeed for a small discrepancy of $u-r\sim 0.1$ at the highest mass end. In light of this, it is nevertheless clear that the X-ray MASSIVE ETGs represent a highly biased sample of red galaxies. This could be due to the fact that the MASSIVE survey probes a volume that is about four times larger than the simulated volume in TNG100. 

Before comparing X-ray quantities between simulations and observations, another stellar quantity that needs to be compared is the effective radius, $R_{\rm e}$, for it is used in X-ray observations to mark the size of the hot atmospheres (e.g. \citealt{goulding.etal.2016,babyk.etal.2018, lakhchaura.etal.2019}). This is of particular importance for massive galaxies, as their hot atmospheres can extend well beyond the stellar distribution and well into the intra-group/cluster medium. For this purpose, we show in the lower panel of Fig. \ref{fig:4} the effective radius-magnitude relations for the simulated and observed samples. In agreement with the findings by \citealt{genel.etal.2017} who compared TNG100 optical sizes to a number of observational results, simulated and observed datasets of X-ray luminous ETGs occupy similar regions in the radius-magnitude space. Marginalizing over selection biases and possible mismatches in the ways the effective radii are operationally measured, this result assures that the properties of the X-ray emitting gas are measured across spatial regions that are consistent within a factor of 1-1.5 between TNG and the observed samples. In fact, the effects of numerical resolution are the reason why TNG100 galaxies have somewhat larger sizes at fixed magnitude than TNG50 ones, by a factor of up to 1.5 or so \citep[see also][]{pillepich.etal.2019}.

\subsection{TNG and observed X-ray relations for ETGs}
\label{sec:comparing}
\begin{figure*}
    \centering
    \includegraphics[width=0.99\textwidth]{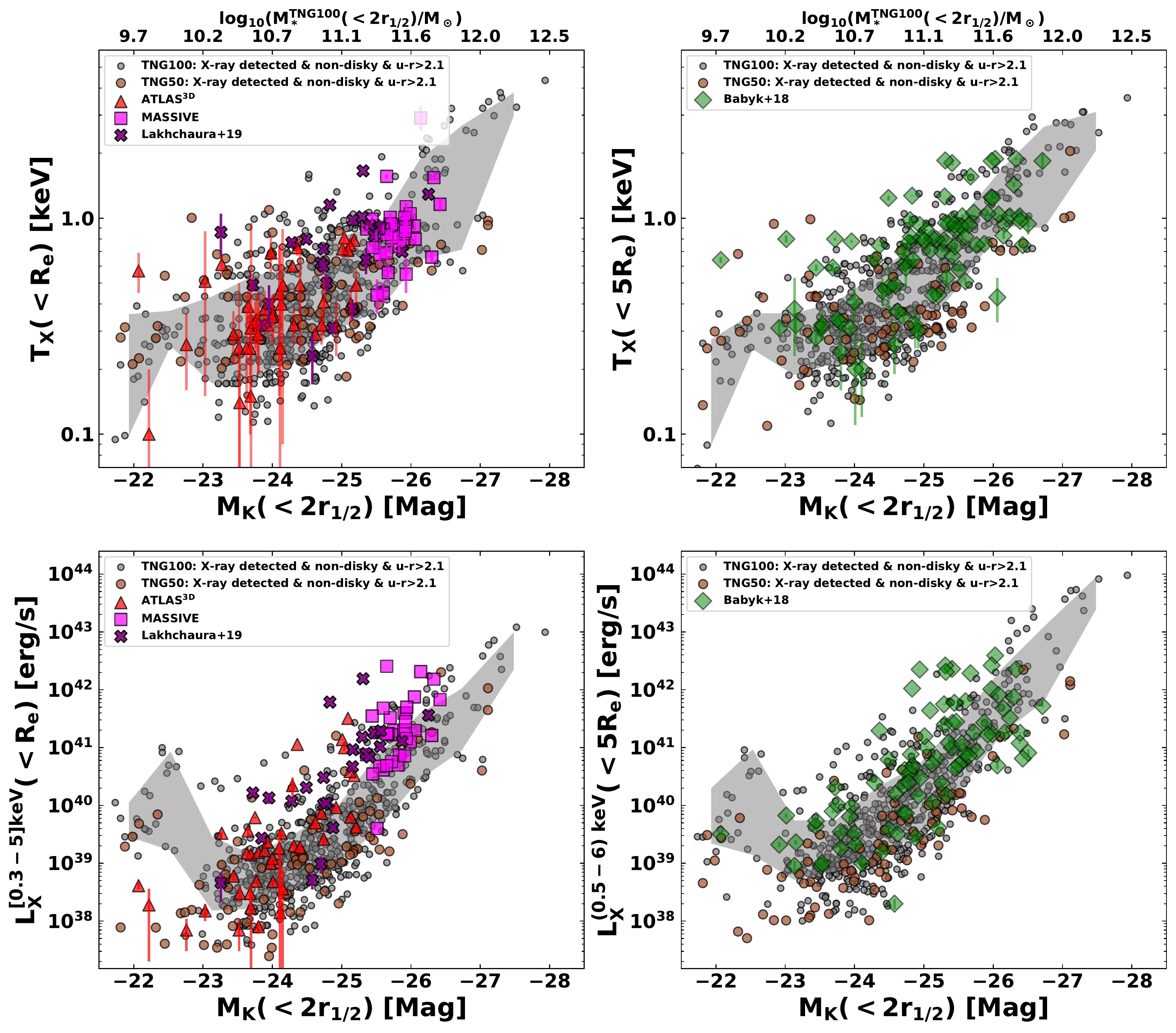}
    \caption{Comparison between simulated and observed ETG X-ray relations measured within $R_{\rm e}$ ({\it left}) and $5R_{\rm e}$ ({\it right}) radius, at $z\simeq0$. The TNG100 and TNG50 selected samples (grey and brown circles) comprise ETG-like galaxies, i.e. non-disky, magnitude-selected ($\rm{M_{\rm K}}<-21.5$), and color-selected ($u-r>2.1$) galaxies in addition to a selection in stellar mass and X-ray measurements (see text in Section \ref{sec:matching}). The shaded area represents $1\sigma$ uncertainty about the median relation for the TNG100 sample.} 
    \label{fig:5}
\end{figure*}
In Fig. \ref{fig:5}, we present the comparison between the simulated and the observed $T_{\rm X}-M_{\rm K}$ and $L_{\rm X}-M_{\rm K}$ relations for ETGs measured within $R_{\rm e}$ (left) and $5R_{\rm e}$ (right). Overall, TNG and the available observations show similar trends of X-ray properties over the considered range of magnitudes: in fact, they all occupy similar regions in the X-ray-magnitude parameter space, returning a non trivial validation of the TNG model and its underlying models for the feedback processes. 

At the bright end of the $M_{\rm K}$ distribution ($M_{\rm K}<-24$), namely for the most massive galaxies ($M_*>5\times10^{10}M_\odot$), both the TNG and the observed X-ray properties (in logarithmic scale) can be reasonably described by a linear function of the magnitude. On the other hand, at the faint end ($M_{\rm K}>-24$), the X-ray temperatures (luminosities) flatten (upturn) for lower-mass galaxies and exhibit larger scatter. The upward tail in the $L_{\rm X}-M_{\rm K}$ relation occurs below $M_{\rm K}\sim-24$, which represents the transition scale between star-forming and quenched galaxies. As we will explicitly demonstrate and discuss later, this reflects the significant variations in the X-ray luminosities of the two galaxy populations.

To quantify the linear dependence on $M_{\rm K}$ at the bright end, we describe and fit the X-ray properties with the following formula:
\begin{equation}
    \log_{10} (F/F_0) = \alpha+\beta\times (M_{\rm K}+26), \label{eqn6}
\end{equation}
where $F$ represents either $T_{\rm X}$ or $L_{\rm X}$, and $F_0$ is the pivotal point. The parameters $\alpha$ and $\beta$ are the best-fitting normalization and slope, respectively. Using the fitting package {\sc linmix\underline{ }err} \citep{kelly.2007}, we fit to equation (\ref{eqn6}) both the TNG100 simulated result, which covers a sufficiently large range of $M_{\rm K}$ for systems with $M_{\rm K}<-24$, as well as the observed data: best-fitting parameters are reported in Table \ref{tb2}.  

 \begin{table*}
  \caption{\label{tb2}
  Best-fitting parameters of equation (\ref{eqn6}) for ETG-like TNG100 galaxies ($M_*>3\times10^9M_\odot$, non-disky, $M_{\rm K}<-24$, $u-r>2.1$, and X-ray detected) and the observed samples at $z\simeq0$. We provide the observational fit for mere reference as in fact it is done on a combination of data points with incongruous selections and measurements. For the TNG fits, note that the X-ray properties are determined via Chandra-like mock observations with 100 ks of exposure time and including the telescope responses. Here TNG galaxy magnitudes account for all the K-band light from within twice the stellar half mass radius.}
 \begin{center}
  \resizebox{0.95\textwidth}{!}{
 \begin{tabular}{cc|cccc|ccc}
 \hline
Relation & $F_0$ & $\alpha$ & $\beta$ & scatter &$\mid$& $\alpha$ & $\beta$ & scatter \\
\hline
 Within $R_{\rm e}$ & & & TNG100   & &$\mid$& & Observations  &  \\
 & & &   & &$\mid$& &($\rm{ATLAS^{3D}}$+MASSIVE+Lakhchaura+19) &  \\
 $T_{\rm X}-M_{\rm K}$ & 1 keV & $-0.03\pm0.02$ &$-0.27\pm0.01$ & $0.21\pm0.01$ &$\mid$& $-0.01\pm0.03$ & $-0.20\pm0.03$ & $0.15\pm0.02$ \\
 $L_{\rm X}^{[0.3-5]\rm{keV}}-M_{\rm K}$ & $10^{40}$ erg/s & $0.77\pm0.05$ &$-1.10\pm0.04$ & $0.56\pm0.02$ &$\mid$& $1.54\pm0.11$ & $-1.10\pm0.12$ & $0.60\pm0.06$ \\
 \hline
 Within $5R_{\rm e}$ & & & TNG100   & &$\mid$& & Observations (Babyk+18) &  \\
 $T_{\rm X}-M_{\rm K}$ & 1 keV & $0.02\pm0.01$ &$-0.29\pm0.01$ & $0.15\pm0.01$ &$\mid$& $0.02\pm0.03$ & $-0.18\pm0.03$ & $0.16\pm0.02$ \\
 $L_{\rm X}^{[0.5-6]\rm{keV}}-M_{\rm K}$ & $10^{40}$ erg/s & $1.54\pm0.05$ &$-1.20\pm0.04$ & $0.61\pm0.02$ &$\mid$& $1.51\pm0.13$ & $-0.83\pm0.15$ & $0.72\pm0.07$\\
 \hline
 \end{tabular}}

 \end{center}
 \end{table*}
The comparison between TNG100 and TNG50 allows us to assess the effects of numerical resolution and sampling. It is already known that within the TNG model improved resolution implies smaller sizes (see previous Section) and larger galaxy stellar masses and luminosities \citep[up to factors of 1.2-1.4 in galaxy stellar mass at $z=0$ in these mass ranges, e.g.][]{pillepich.etal.2018}. Nevertheless, overall TNG100 and TNG50 galaxies occupy similar regions in the parameter spaces with two noticeable differences. At the highest-mass end, TNG50 X-ray temperatures appear biased low compared to TNG100: this is probably due to the absence of a large number of galaxies living at the center of massive clusters, as the TNG50 volume is almost 10 times smaller than TNG100 and the most massive TNG50 haloes have masses of $10^{14}M_\odot$. Moreover, on average, the TNG50 X-ray temperatures and luminosities within $5R_{\rm e}$ at fixed magnitude are smaller by a up to a factor of a few: this could be due to the effects of resolution on the nominal apertures for the measurements (smaller in TNG50 than in TNG100) or due to the resulting slightly different thermodynamical properties of the gas, or a combination of both.

Within these uncertainties, the simulated $T_{\rm{X}}-M_{\rm K}$ relations are in good agreement with the observed relations at both radii. There is no significant variation in the gas temperature relations between the two radii, except for a somewhat smaller simulated scatter for larger apertures: 0.21 vs. 0.15 dex for TNG100. For $M_{\rm K}<-24$, the TNG100 simulated $T_{\rm X}-M_{\rm K}$ relation is well approximated by a linear function as described in equation (\ref{eqn6}) with the slope $\beta\sim-0.3$\footnote{We note that $M_{\rm K}$ is negative hence a negative slope means a positive correlation between the two quantities.}. The observed slope is slightly shallower ($\beta\sim-0.2$). As $M_{\rm K}$ is closely related to galaxy stellar mass (as shown in Section~\ref{sec:tng}), these findings imply that the gas temperature, which represents its thermal energy, is primarily determined by a galaxy's potential (see also \citealt{goulding.etal.2016}). At the faint end of the $M_{\rm K}$ distribution, on the other hand, $T_{\rm X}$ starts to stabilize around 0.3 keV yet with larger scatter, being consistent with the idea that lower mass galaxies are more sensitive to non-gravitational heating processes such as stellar and AGN feedback. Even though the energy range for fitting is limited to  the $[0.3-7.0]$ keV band, where {\it Chandra} is sensitive and reasonably well calibrated, it is still possible to detect cool systems with atmospheric effective temperatures down to $T_{\rm X}\sim0.1$ keV, as such gas produces X-ray line emission of OVII and OVIII in the {\it Chandra} band. These lines are strong and their ratios will provide a good ``thermometer'' even if we only detect a relatively small number of photons. 

 For the $L_{\rm X}-M_{\rm K}$ relation, the TNG100 and TNG50 simulations reproduce reasonably well the observed trends at both radii, even though there is a slight offset in normalization (at $<R_{\rm e}$) depending on exactly which simulation and observational datasets are considered. For instance, the median value of the simulated $L_{\rm X}(<R_{\rm e})$, at $M_{\rm K}=-26$, is lower than the observed median value by a factor of $\sim5$. Yet, given the large scatter in both simulated and observed samples, the discrepancy is less than $1\sigma$. At larger radii, $<5R_{\rm e}$, the ratio between observed (\citealt{babyk.etal.2018}) and TNG100 $L_{\rm X}$ median values is less than a factor of 2 and they are fully statistically consistent. For TNG100, the X-ray luminosity measured within $5R_{\rm e}$ is increased by a factor of $\sim5$ on average with little dependence on magnitude with respect to the one measured within $R_{\rm e}$. 
 
 Taken at face value, given that the simulations and all the observational datasets appear to be consistent for the $T_{\rm X}-M_{\rm K}$ relations, the difference in X-ray luminosity could be an indication of a lack of hot gas within the central regions of simulated galaxies compared to observations. For instance, given that $L_{\rm X}\propto f_{\rm g}^2$, where $f_{\rm g}\equiv M_{\rm gas}/M_{\rm tot}$ is the hot gas fraction, an offset in the X-ray luminosity by a factor of $\sim5$ can be translated into an offset in the hot gas fraction by a factor of $\sim2$. The result might in turn suggest a too strong black hole feedback in the most massive galaxies which blows out too much gas from the central regions. In fact, the mismatch could be more simply due to a difference (actual or of measurement) in the K-band luminosity towards the highest-mass end. In practice, the small offset might be of observational origin. As discussed in Section~\ref{sec:props}, the observed $M_{\rm K}$ could be underestimated due to the relative shallow photometry of the 2MASS survey. For illustration, if we use $M_{\rm K}(<30\ {\rm kpc})$ instead of $M_{\rm K}(<2\times r_{1/2})$, the $L_{\rm X}$ offset between simulations and observations is reduced to a factor of $\sim4$. Finally, we note that the offset in $L_{\rm X}$ is mainly caused by the high-mass end galaxies of the MASSIVE sample, which is obtained from a $\sim4$ times larger volume compared to the simulated volume (TNG100), and it could be biased toward the most X-ray luminous galaxies in the nearby Universe.    
 
Another point worth noting is that the scatter in $L_{\rm X}$ at a given magnitude is remarkably larger, about 3 times, than the temperature scatter in both simulated and observed samples. This suggests that for massive galaxies at the same stellar mass, their X-ray luminosity may depend significantly on other galaxy properties that affect or correlate with the amount of hot gas, such as galaxy kinematics (e.g. fast versus slow rotators, see \citealt{sarzi.etal.2013}). Further investigation into the scatter of the X-ray luminosity, though intriguing, is beyond the scope of the current work and should be addressed in a detailed future study. 
\begin{figure*}
    \centering
    \includegraphics[width=0.89\textwidth]{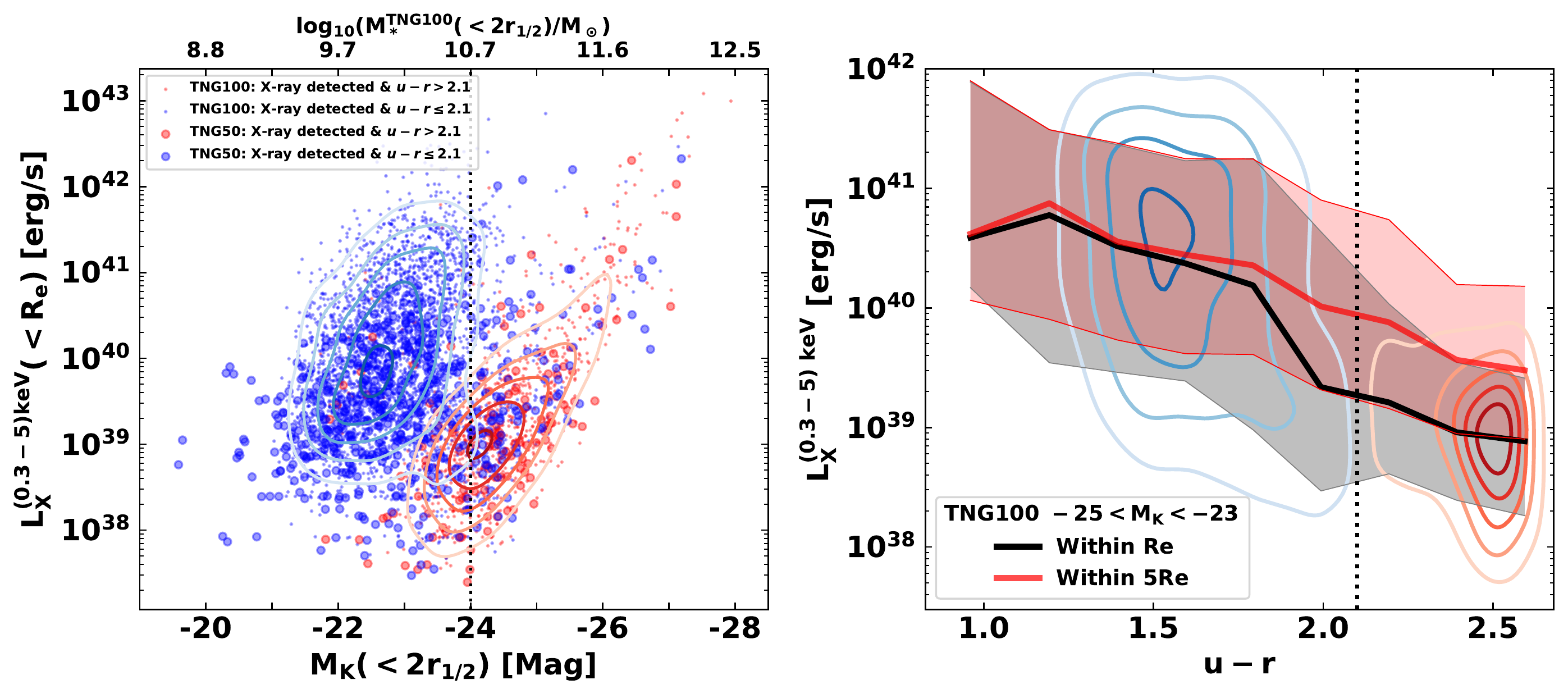}
    \includegraphics[width=0.89\textwidth]{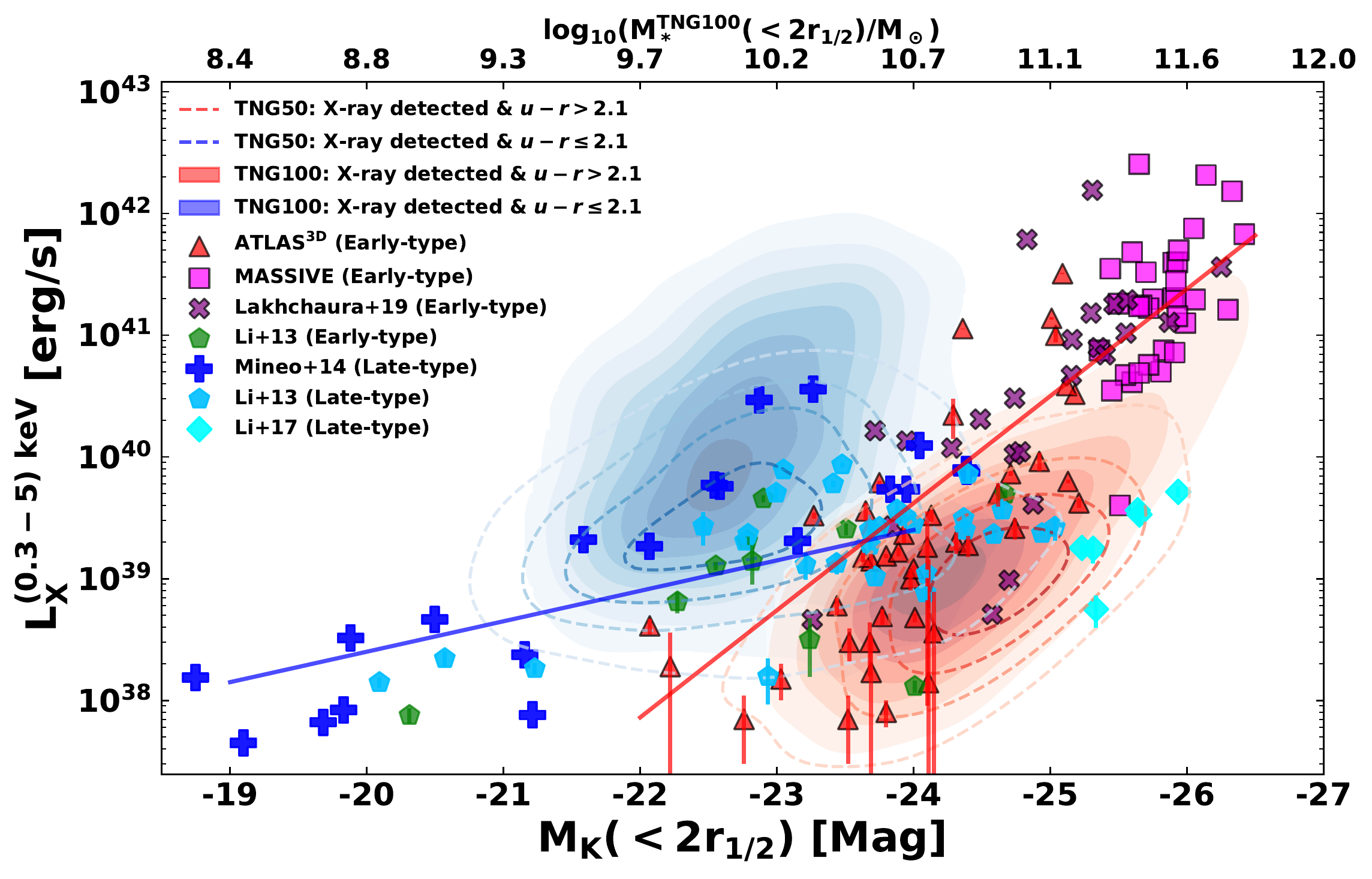}
    \caption{ The dependence on galaxy type of the X-ray luminosity from the diffuse, hot and metal-enriched gas. {\it Top row:} the left panel shows the TNG $L_{\rm X}-M_{\rm K}$ relation within $R_{\rm e}$ in which the data points are color-coded according to their $u-r$ information with respect to the threshold value that defines ETGs as presented in Sections~\ref{sec:props} and \ref{sec:matching}. The contours locate the blue cloud and red sequence loci of the TNG100 simulation. In the right panel, a relation between $L_{\rm X}$ and $u-r$ is shown for TNG100 systems with $-25<M_{\rm K}<-23$ at two different radii: $R_{\rm e}$ and $5R_{\rm e}$. The contours specify the loci of the blue cloud and the red sequence for the case where $L_{\rm X}$ is measured within $R_{\rm e}$. {\it Bottom row:} X-ray luminosity as a function of $M_{\rm K}$ magnitude is shown for TNG simulated galaxies in comparison with X-ray observations. The TNG100 and TNG50 data is represented by color-filled and dashed-line contours, respectively, color-coded according to their $u-r$ colors. The two blue and red solid lines represent the best-fit relations, as reported in equations (\ref{eqn7}) and (\ref{eqn8}), for the observed late- and early-type galaxies, respectively.}
    \label{fig:6}
\end{figure*}
\section{The dependence of X-ray properties on galaxy type}
\label{sec:predictions}
After verifying that the TNG simulations realistically reproduce the observed X-ray properties of the hot atmospheres in massive ETGs, we now use the TNG simulations to get insights on the X-ray properties of low-mass and star-forming galaxies. Of particular interest is how the X-ray properties depend on galaxy type. This follows up on previous studies from observational works (e.g \citealt{mineo.etal.2012, li.wang.2013b, li.etal.2017}), which reveal a tight correlation between galaxies gaseous X-ray emission and their star-formation rates, whereby suggesting a close connection between the hot diffuse gas content and the galaxies star-formation state. This result is tentatively supported by numerical studies (e.g. \citealt{crain.etal.2010,davies.etal.2019b}): for instance, \cite{davies.etal.2019b} show that, using the EAGLE simulation, the star-formation rate of simulated galaxies are significantly positively correlated with their gas content, which is in turn manifested in observable quantities such as the X-ray luminosity or the thermal Sunyaev-Zel'dovich flux integrated throughout the halo. \citealt{crain.etal.2010} found similar correlations with earlier numerical simulations of galaxies.
%
\subsection{The $L_{\rm X}-M_{\rm K}$ relations for star-forming and quenched galaxies}
We first examine the simulated $L_{\rm X}-M_{\rm K}$ relation in connection to the star formation activity of the galaxies using their color, i.e. $u-r$, as a proxy for their star formation state. 

The top left panel in Fig. \ref{fig:6} shows the $L_{\rm X}-M_{\rm K}$ relation in TNG100 and TNG50 in which the data points are color-coded according to their $u-r$ values. We apply the same color cut as presented in Section~\ref{sec:props}, i.e. $u-r>2.1$, to define quenched or red galaxies (colored in red), and $u-r\leq2.1$ to select the star-forming galaxies (colored in blue). The number fraction of quenched (star-forming) galaxies is about $20\%$ ($80\%$) and $18\%$ ($82\%$) for TNG100 and TNG50, respectively. Density contours are drawn to indicate the most populated regions of the two populations in TNG100. A clear pattern emerges in both TNG100 and TNG50, in which the quenched galaxies occupy the bright end ($M_{\rm K}<-24$) but are clustered around a relatively low $L_{\rm X}$ ($L_{\rm X}\sim10^{39}$ erg~s$^{-1}$), while the star-forming galaxies populate the faint end ($M_{\rm K}>-24$) and their $L_{\rm X}$ is centered around a higher value ($L_{\rm X}\sim10^{40}$ erg~s$^{-1}$). Critically, at the magnitude range where the star-forming and quenched galaxies overlap ($M_{\rm K}\sim-24$), the TNG simulations predict a clear X-ray luminosity separation between the two populations, with the star-forming galaxies being X-ray {\it brighter} than the quenched systems. The separation is more pronounced in TNG100 than in TNG50. 

To better quantify how the X-ray luminosity depends on the star-formation state, we select simulated galaxies in TNG100 whose $M_{\rm K}$ falls in the range where the star-forming and quenched galaxies overlap, i.e. $-25<M_{\rm K}<-23$ within which the quenched (star-forming) fraction is about $30\%$ ($70\%$), and plot their median values of $L_{\rm X}$ as well as the corresponding $1\sigma$ scatter as a function of $u-r$, as shown in the top right panel in Fig. \ref{fig:6}. We show the results for $L_{\rm X}$ measured within two apertures, $R_{\rm e}$ and $5R_{\rm e}$, to explore how this effect depends on radius. At both apertures, the X-ray luminosity decreases steeply with color, falling by more than one order of magnitude from the blue to the red end. At the red end, the luminosity measured at an intermediate radius, i.e. within $5R_{\rm e}$, is systematically larger than the one measured within $R_{\rm e}$. This result suggests that in the range of K-band magnitudes where the two populations overlap, the difference in $L_{\rm X}$ between the star-forming and quenched galaxies is more prominent when measured in the central regions than at larger radii.    

The dichotomy in X-ray luminosity between star-forming and quenched galaxies is intriguing since it is an observationally testable prediction. 
In addition to the observations of early-type galaxies presented in Section~\ref{sec:comparing}, we also compare our results with {\it Chandra} and {\it XMM-Newton} observations of lower-mass late-type galaxies taken from \cite{mineo.etal.2012}, \cite{li.wang.2013}, and \cite{li.etal.2017}.
 The final observed sample consists of 163 galaxies (108 early-type, 55 late-type) spanning a range of over 7 magnitudes ($-26<M_{\rm K}<-19$).    

For comparison with X-ray observations of disk galaxies, it is worth mentioning a caveat regarding the difference in studied volume of the hot atmospheres between simulations and observations. Unlike the case of massive elliptical/lenticular galaxies, the X-ray observations of disk galaxies (e.g. \citealt{mineo.etal.2012, li.wang.2013}) are mainly taken from a boxy volume with the size characterized by $D25$, which is defined as the B-band projected diameter of the ellipse major axis at isophotal level $25\ \rm{mag}\ \rm{arcsec}^{-2}$, and $r25$ which is the ratio of the major to minor axes of the ellipse (see, e.g. Fig. 4 in \citealt{li.wang.2013} for an illustration). Instead, in this work we compute the simulated X-ray quantities within a cylindrical volume characterized by the effective radius for both star-forming and quenched galaxies, as described in Section~\ref{sec:mocks}. Compared to simulations (not shown here), the available observed values, taken from \cite{li.wang.2013}, of the major and minor axes of the D25 ellipse fall within the range of the simulated $R_{\rm e}$ distribution. Moreover, we  verify that for low-mass systems, the simulated X-ray measurements within $R_{\rm e}$ cover the bulk of the total galactic X-ray emission of that galaxy, e.g. for galaxies with $M_{\rm K}>-24$, $L_{\rm X}(<R_{\rm e})$ contributes more than $80\%$ of the total luminosity ($[L_{\rm X}(<5R_{\rm e})$). This result justifies the use of the simulated measurements within $R_{\rm e}$ for the comparison with the late-type observations.     

In the bottom panel of Fig. \ref{fig:6}, we show the X-ray luminosity-magnitude relation for both observed late- and early-type galaxies (datapoints in shades of blue and red, respectively) overplotted to TNG simulated data represented by contours that locate the loci of star-forming (blue) and quenched (red) galaxies in TNG100 (color-filled) and TNG50 (dashed-line contours). 

Neglecting the inhomogeneity of the datasets collected here, the observed $L_{\rm X}-M_{\rm K}$ relation can be approximately described by a broken linear function of magnitude, in which early-type and late-type galaxies follow two distinct linear relations:

\begin{equation}
\resizebox{0.49\textwidth}{!}{
    ${\rm Early-type:}\ \log_{10}\big(\frac{L_{\rm X}}{10^{40}{\rm erg/s}}\big)=(-21.5\pm1.3)+(-0.88\pm0.05)\times M_{\rm K}, \label{eqn7}$ 
    }
\end{equation}
\begin{equation}
\resizebox{0.49\textwidth}{!}{
    ${\rm Late-type:}\ \log_{10}\big(\frac{L_{\rm X}}{10^{40}{\rm erg/s}}\big)=(-6.6\pm0.9)+(-0.25\pm0.04)\times M_{\rm K}, \label{eqn8}$
    }
\end{equation}
with intrinsic scatters of $0.66$ dex and $0.48$ dex for early-types and late-types, respectively. The early-type relation is significantly steeper, where the slope is larger by about a factor of 3 than the late-type slope. Interestingly, at the joint between the two populations, $M_{\rm K}\sim[-23, -24]$, late- and early-type galaxies are segregated into high- and low-$L_{\rm X}$ regions, respectively, on the $L_{\rm X}-M_{\rm K}$ plane. This segregation causes the scatter in the X-ray luminosity to be remarkably large, with data points scattering over more than two orders of magnitude in $L_{\rm X}$.  
\begin{figure*}
    \centering
    \includegraphics[width=0.99\textwidth]{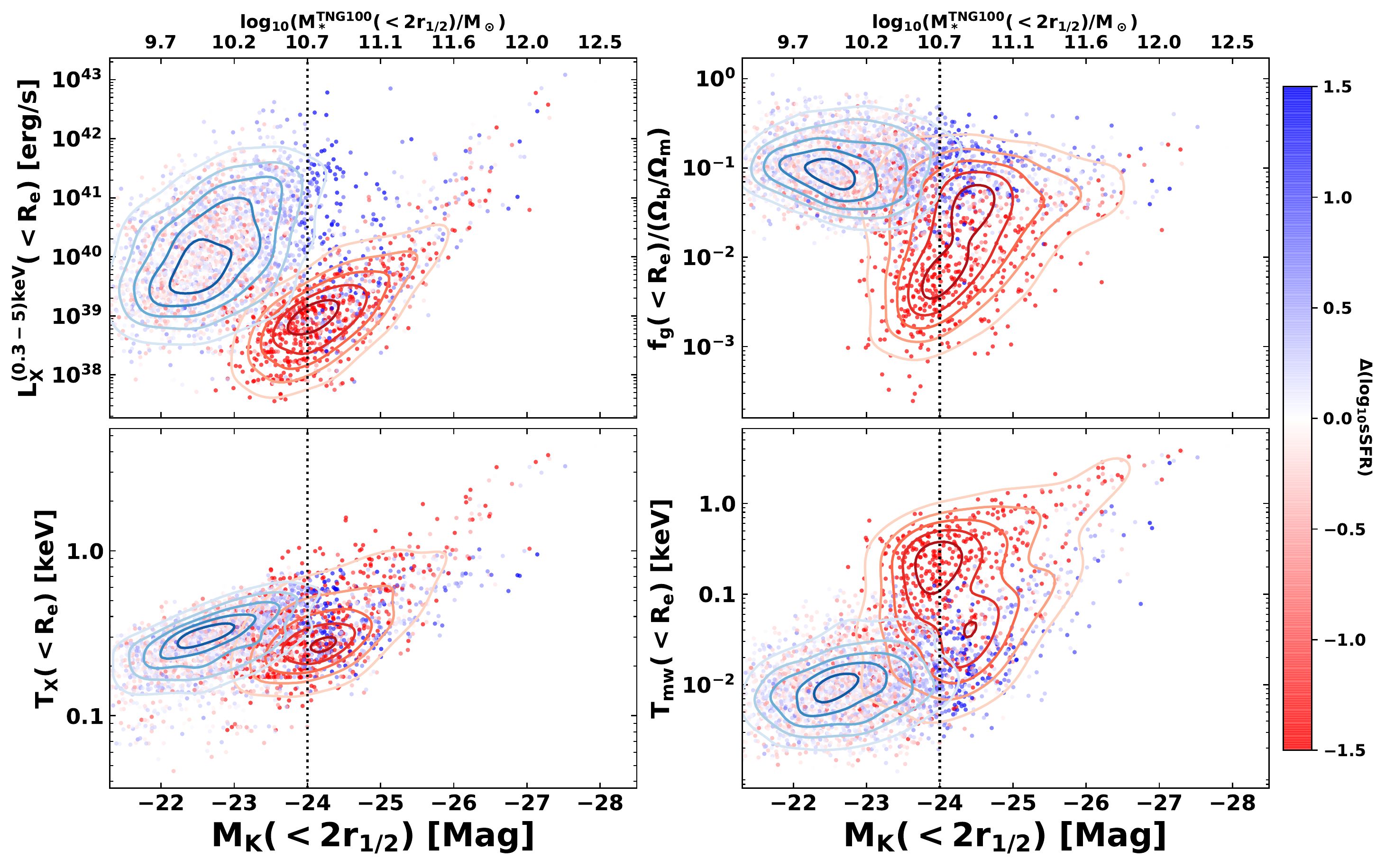}
    \caption{The hot gas properties with respect to the star formation activity in TNG100 at $z=0$. Clock-wise, the plots are shown for $L_{\rm X}-M_{\rm K}$, $f_{\rm g}-M_{\rm K}$ ($f_{\rm g}$ is  normalised to the baryon fraction $\Omega_{\rm b}/\Omega_{\rm m}$ and it is the mass of all gas, not just the hot gas, divided by  the total mass), $T_{\rm mw}-M_{\rm K}$, and $T_{\rm X}-M_{\rm K}$ relations. Note the difference in the plotted temperature range in the y-axis of the two temperature plots. The data points are color-coded according to the relative difference between their specific star formation rate ($\rm{sSFR}$) and the running median of the $\rm{sSFR}-M_{\rm K}$ relation at their magnitude. The vertical dotted line marks roughly the transition point, $M_{\rm K}=-24$, between star-forming and quenched galaxies. The contours specify the locations of the two populations, star-forming (blue) and quenched (red), defined according to the star-formation flags described in Section~\ref{sec:props}.} 
    \label{fig:7}
\end{figure*}

Qualitatively, the tentatively observed segregation of X-ray luminosity between late- and early-type galaxies is consistent with the predictions of the TNG simulations. Nonetheless, the qualitative agreement between observations and simulations at this stage should not be overinterpreted for two reasons: i) the observed sample of late-type galaxies is statistically limited and not well defined in terms of the range in near-infrared magnitude and the probed volume; and ii) observationally it is challenging to obtain the truly diffuse X-ray emission of the hot ISM due to contamination by unresolved emission of compact sources (e.g. high-mass X-ray binaries), whose contribution can only be estimated in a statistical and model-dependent way (see e.g. \citealt{mineo.etal.2012} for a thorough study). Another point worth discussing here is the fact that in the magnitude range $-24<M_{\rm K}<-22$, we see a number of remarkably luminous late-type galaxies ($\sim16\%$) with $L_{\rm X}\gtrsim10^{41}$ erg~s$^{-1}$ in TNG100, which are as bright as the most massive giant ellipticals. Such highly X-ray luminous atmospheres in late-type galaxies have so far not been observed in the local Universe, which may imply that the TNG simulations overpredict the gas phase X-ray emission for a fraction of star-forming galaxies. This in turn may link to the inefficiency of the implemented feedback models in removing gas out of the central regions thereby resulting in a number of too gas-rich galaxies at the low-mass end. Or the overpredicted X-ray luminosity could result from that the gas is over-heated by both AGN and stellar feedback in galaxies at the transition range (see discussion at the end of Section \ref{sec:5}). More detailed analyses of those highly-luminous star-forming galaxies as well as the connection with the associated feedback processes, which are beyond the scope of the current study, would provide essential diagnostics for the TNG model. Besides, the discrepancy between TNG and observations could be partly due to the fact that there has not been a sensitive high spatial resolution all sky X-ray surveys of star-forming galaxies and TNG100 encompasses a larger volume, about $5$ times, than the one probed by the current late-type observations. The upcoming all sky survey with {\it eROSITA} \citep{merloni.etal.2012} will provide excellent X-ray catalogs of nearby galaxies to examine this prediction of the TNG model. 
\section{The origin of the Luminosity Diversity}
\label{sec:5}
The diversity in $L_{\rm X}$ presented in the previous Section suggests that the X-ray emission of hot atmospheres is closely linked to the star formation state of the host galaxies. In turn, at least within the TNG galaxy formation model, this is closely connected to the stellar and SMBH feedback processes. To further investigate the origin of the X-ray luminosity difference in star-forming and quenched galaxies, in this Section we inspect the thermal properties as well as the gas content in the two sets of systems and their connection with the SMBH driven feedback. For the sake of brevity, we only employ the TNG100 sample for this theoretical study, though the following results are also qualitatively applicable to the TNG50 sample.

\subsection{Connection with Gas content}
In Fig. \ref{fig:7} we show the TNG100 relations between the X-ray luminosity, gas mass fraction ($f_{\rm g}$), X-ray temperature ($T_{\rm X}$) and mass-weighted temperature ($T_{\rm mw}$) measured within $R_{\rm e}$ as a function of the K-band magnitude, in which the data points are color-coded based on their specific star formation rate (${\rm sSFR}$) with respect to the median of the ${\rm sSFR}-M_{\rm K}$ relation at the same magnitude. In this way, we can isolate the intrinsic correlation between X-ray properties or gas content and sSFR from their correlation with the K-band magnitude that traces the stellar mass.
On top of that, contours are used to indicate the parameter space occupied by star-forming (blue) and quenched (red) galaxies, which are flagged depending on their relative position with respect to the main sequence (see Section~\ref{sec:props} for a detailed definition). Analog plots for the gas density and metallicity can be found in Appendix \ref{sec:appB}.

As expected, the two populations of quenched and star-forming galaxies, when flagged according to their instantaneous star formation rate, display a similar separation on the $L_{\rm X}-M_{\rm K}$ plane as in the case of the color-based study presented in Section~\ref{sec:predictions}. However, when color-coded by $\rm{sSFR}$ at a given magnitude, there is no clear segregation between galaxies above and below the median $\rm{sSFR}$ value across the considered magnitude range except at $M_{\rm K}\sim-24$. At this magnitude (stellar mass), highly star-forming galaxies are much more X-ray luminous than their counterparts with low star-formation rates. Importantly, the gas mass fraction is also found to have the largest scatter at this magnitude, where galaxies below the median $\rm{sSFR}$ are gas depleted by more than an order of magnitude, compared to those above the median. 

The temperature relations, $T_{\rm X}-M_{\rm K}$ and $T_{\rm mw}-M_{\rm K}$ display different patterns for star-forming versus quenched galaxies. The former shows no clear division between the two populations even at $M_{\rm K}\sim-24$, where star-forming and quenched galaxies have similar X-ray temperatures ($T_{\rm X}\sim0.2-0.3$ keV). On the other hand, a clear separation is found in the $T_{\rm mw}-M_{\rm K}$ plane where the quenched galaxies are about an order of magnitude hotter than their star-forming counterparts. The different patterns between the two temperature estimators can be explained by the fact that the X-ray temperatures are mainly determined by the temperatures of the gas cells that emit efficiently in the X-ray band (i.e. $[0.3-5]$ keV)\footnote{In fact, $T_{\rm X}$ is a close estimator of the emission-weighted temperature ($T_{\rm ew}$). See Appendix~\ref{sec:appA} for a detailed comparison of temperature estimators.}, therefore it is likely biased high, especially for low-mass systems, compared to the mass-weighted temperature, which presumably represents the averaged thermal energy of all the considered gas cells.      

\begin{figure*}
    \centering
    \includegraphics[width=0.79\textwidth]{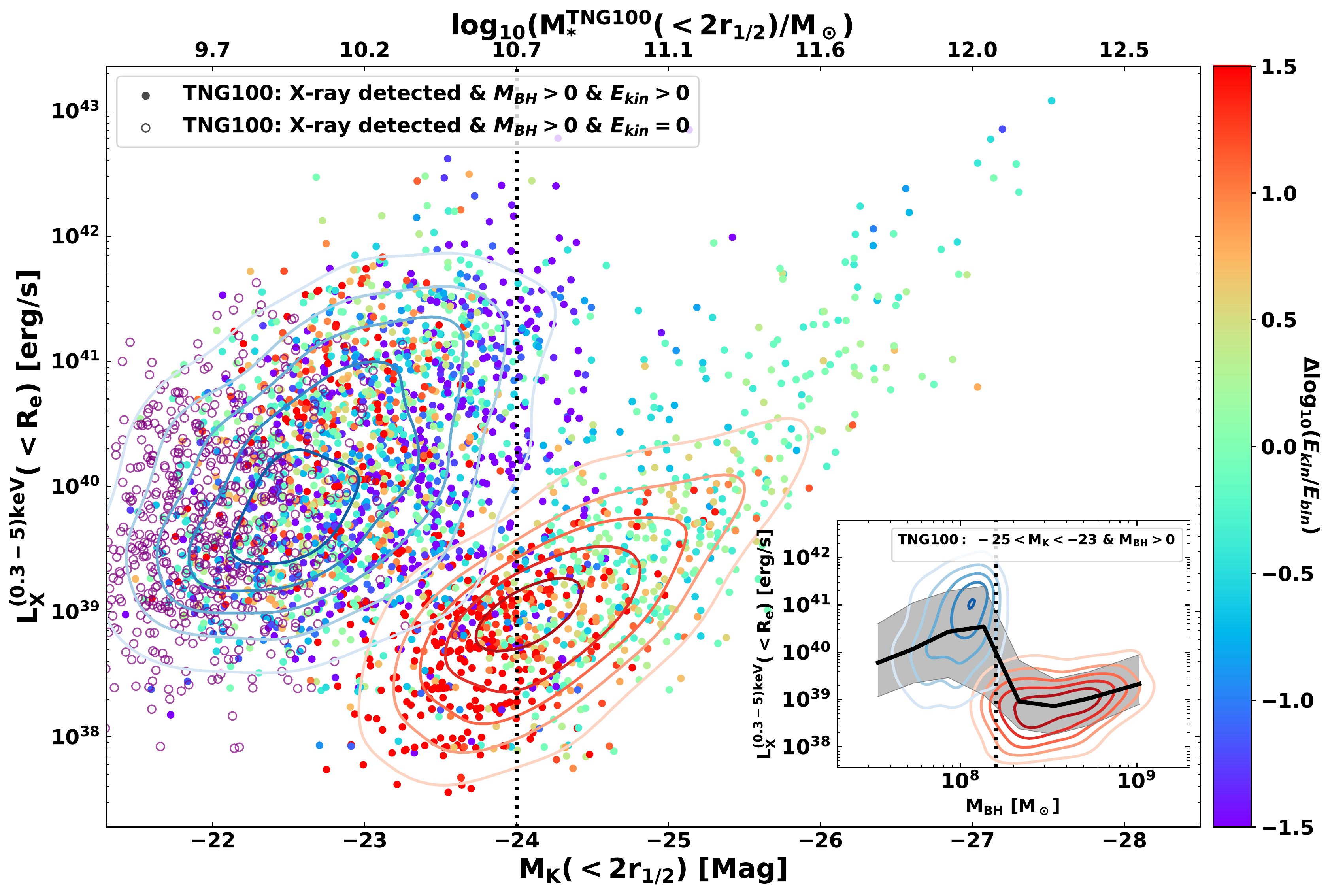}
    \caption{Main panel: the effects of SMBH kinetic feedback on X-ray luminosity across the explored magnitude range for TNG100. The data points are color-coded according to the relative difference between their value of feedback-to-binding energy ratio ($E_{\rm kin}/E_{\rm bin}$, see the text for definition) and the running median of the $E_{\rm kin}/E_{\rm bin}-M_{\rm K}$ relation at their magnitude. This energy ratio reflects the capacity of SMBH feedback to push gas out of the central region of the galaxy. It becomes clearly visible at $M_{\rm{K}}\sim-24$ ($M_*\sim10^{10.7}M_\odot$, dotted line), where the SMBH feedback becomes sufficiently powerful to push out large amounts of hot gas thereby significantly reducing the X-ray luminosity. The contours represent the two populations of star-forming and quenched galaxies, identical to those in the top-left panel of Fig. \ref{fig:7}. For completeness, we also show galaxies with $E_{\rm kin}=0$ as empty circles, namely those galaxies in which SMBH feedback is still in the thermal mode. Sub panel: the $L_{\rm X}-M_{\rm BH}$ relation for TNG100 galaxies with magnitude in the transition range ($[-25, -23]$) is represented by the median relation and the $1\sigma$ envelope. The contours locate the loci of star-forming (blue) and quenched (red) populations which are separated at $M_{\rm BH}\sim10^{8.1}M_\odot$ (denoted by the vertical dotted line).}
    \label{fig:8}
\end{figure*}

In summary, the results shown in Fig. \ref{fig:7} reveal that the quenched galaxies in TNG are on average hotter but poorer in gas than the star-forming systems. We further note that the quenched population is also slightly metal-poorer than the star-forming (see Appendix~\ref{sec:appB}). Though overall the contribution from metal line emission dominates the total gaseous X-ray emission of galaxies across the whole considered range of magnitude, we notice that being metal-poorer is not the main reason for which quenched galaxies exhibit significantly less X-ray luminosity than the star-forming population: we demonstrate and discuss this in some detail in Appendix \ref{sec:appB}. All these results indicate that gas depletion is primarily responsible for the lower X-ray luminosity of quenched galaxies. In other words, quenched galaxies have lower values of $L_{\rm X}$ because they contain less (albeit hotter) gas than star-forming galaxies. This finding explains the previous connection between the diversity in $L_{\rm X}$ and galaxy types and is consistent with previous results (e.g. \citealt{nelson.etal.2018, nelson.etal.2018b,terrazas.etal.2019, davies.etal.2019}) which indicate that gas removal is the primary cause of star formation quenching in TNG galaxies, in addition to gas heating (Zinger et al. in prep.).  
\subsection{Connection to black hole feedback}
The results found in the previous Section provide an important connection between the gas content, which is closely linked to the quenching mechanism, and the diversity in $L_{\rm X}$. Since the latter can be verified observationally, it offers a critical test for the quenching mechanism in the TNG simulations. 

Previous studies of BH feedback in TNG, e.g. \cite{weinberger.etal.2017}, \cite{weinberger.etal.2018}, \cite{nelson.etal.2018}, and \cite{terrazas.etal.2019}, suggest that the kinetic mode of SMBH feedback may play an important role in quenching star formation. The feedback does not only heat the gas, but it can also lift gas to higher altitudes or strip the galaxy of its star-forming material. Here, we aim to explore the imprint of the SMBH kinetic feedback on the X-ray luminosity of the hot atmospheres. For this task, we consider galaxies that host at least one supermassive black hole at their center and which have already switched to the kinetic mode (i.e. $\int \dot{E}_{\rm kin}dt>0$). By following \cite{terrazas.etal.2019}, we compute the ratio of the accumulated SMBH kinetic feedback to the gas binding energy $E_{\rm kin}/E_{\rm bin}$, as described in Section~\ref{sec:props}.

In the main panel of Fig. \ref{fig:8}, we show the $L_{\rm X}-M_{\rm K}$ relation for TNG100 galaxies, color-coding the data points according to the relative difference with respect to the median value of the energy ratio $E_{\rm kin}/E_{\rm bin}-M_{\rm K}$ relation at the given magnitude. For completeness, the galaxies in which SMBH feedback is still in the thermal mode ($E_{\rm kin}=0$) are also presented (empty circles). In addition. we specify the loci of star-forming and quenched galaxies via contours as done in the top-left panel of Fig. \ref{fig:7}. 

In general, the energy ratio does not show much scatter, except in the range of $M_{\rm K}\sim[-23, -25]$, where the ratio varies by up to three orders of magnitude. As shown in previous studies (e.g. \citealt{terrazas.etal.2019}), for low-mass systems ($M_*<10^{10.7}M_\odot$), the gas binding energy appears larger than the accumulated BH kinetic feedback. As galaxies increase their mass to $M_*\sim10^{10.7}M_\odot$ ($M_{\rm K}\sim-24$), at which scale the mass of the central SMBH reaches the critical value $M_{\rm BH}\simeq10^8M_\odot$ as described in equation (\ref{eqn1}), the integrated kinetic feedback energy ($E_{\rm kin}\simeq10^{59}$ erg) starts to dominate the galactic gravitational potential, causing the energy ratio to increase by over three orders of magnitude.

We note that at $M_{\rm K}\sim-24$, where the energy ratio exhibits the largest scatter, we see a clear separation in X-ray luminosity for galaxies above and below the median value of $E_{\rm kin}/E_{\rm bin}$. Galaxies above the median value, namely systems where the kinetic feedback dominates the binding energy, are significantly fainter in X-rays compared to the galaxies that lie below the median. This result clearly suggests a casual relationship between the SMBH driven feedback activity and the diversity in $L_{\rm X}$, where the SMBH kinetic feedback lifts appreciable amounts of gas from the central potential of massive galaxies, driving their X-ray luminosity low, and quenching their star formation. It also explains the origin of the mass scale, $M_*\sim10^{10.7}M_\odot$ ($M_{\rm K}\sim-24$), where the $L_{\rm X}$ diversity occurs, as it corresponds to the scale where SMBHs start to effectively switch from thermal to kinetic mode of feedback \citep[see][ for a discussion]{weinberger.etal.2017, terrazas.etal.2019, davies.etal.2019}.

As in the TNG model the SMBHs feedback activity is closely connected to their mass, to illustrate this point we show in the sub panel of Fig.~\ref{fig:8} the relation between X-ray luminosity and BH mass for a subsample of TNG100 galaxies that lie in the transition range, i.e. $-25<M_{\rm K}<-23$. As shown in the plot, the two populations are separated at $M_{\rm BH}\sim10^{8.2-8.3}M_\odot$ and a remarkable drop in the X-ray luminosity from star-forming to quenched galaxies, in which the latter is about an order of magnitude less luminous than the former. The sharp division between the two populations at $M_{\rm BH}\sim 10^{8.2-8.3}M_\odot$, which result from an ensemble of choices in the model in addition to the Eddington ratio threshold for switching SMBH feedback from thermal to kinetic mode as described in equation (\ref{eqn1}), has been challenged by some observational data (see \citealt{terrazas.etal.2019} for a detailed discussion) which prefer a broader scatter in $M_{\rm BH}$ at the transition range than what emerges from the TNG simulations.

Our finding is in line with recent theoretical studies by \citealt{davies.etal.2019b, davies.etal.2019} based on the EAGLE simulations: according to the EAGLE galaxy formation model, a strong negative correlation is found between the scatter in the gas content -- which can be probed via X-ray and Sunyaev-Zel'dovich (SZ) observations (see Figure 4 in \citealt{davies.etal.2019b}) -- and the scatter in the SMBH mass at a fixed halo mass. The result is in turn linked to the ability of the SMBHs to expel gas via feedback, which also in the EAGLE simulations play an essential role in quenching star formation in central galaxies. Despite the differences in the implemented model of SMBH feedback in EAGLE compared to the TNG model (see \citealt{davies.etal.2019} for a full discussion), the results agree on the crucial role played by SMBH feedback on establishing the population of quenched galaxies via gas ejection, in addition to gas heating.

Finally, SMBH kinetic feedback might not be solely responsible for the observed large separation between the quenched and star-forming galaxies in $L_{\rm X}$. This separation could be further amplified by SMBH thermal and stellar feedback, which could boost the atmospheric X-ray luminosity in star-forming galaxies. For low-mass galaxies ($M_*\lesssim10^{9.5}M_\odot$) in TNG, the SMBH thermal feedback at high-accretion rates is negligible as the injected thermal energy is quickly lost in the star-forming gas phase (see Figure 1 in \citealt{weinberger.etal.2018}). It becomes an important heating channel in the mass range $M_*\sim[10^{10}-10^{10.5}]M_\odot$, which is close to the transition range ($M_{\rm K}\sim-24$), whereby it contributes significantly to the gas phase X-ray emission of star-forming galaxies. In addition, stellar feedback, though being a sub-dominant energy channel at the considered mass range, still releases non-negligible feedback energy and it is expected to both return an appreciable amount of material into the interstellar medium and heat the gas above its virial temperature. 
\section{Summary and conclusions}
\label{sec:6}

In this paper we have investigated the properties of galactic hot atmospheres using a large sample of simulated galaxies taken from the IllustrisTNG cosmological simulations. Specifically, we used galaxies from the TNG100 and TNG50 runs with stellar masses (K-band magnitudes) spanning the $10^{8-12.5} M_\odot$ ($[-17,-28]$) range at $z=0$. We have carried out mock X-ray analyses from the diffuse, hot and metal-enriched gas of the simulated objects as if they were observed with {\it Chandra} and then compared the simulated X-ray scaling relations, such as $T_{\rm X}-M_{\rm K}$ and $L_{\rm X}-M_{\rm K}$, to those obtained from a collection of X-ray observations of nearby galaxies, including for example $\ATLAS$ and MASSIVE early-type galaxies with X-ray measurements. We thus used the simulations to gain critical insights into the diversity of the hot atmospheres and the connection between the kinetic SMBH feedback -- which provides a mechanism for quenching the star formation in the TNG simulations -- and the X-ray properties of the gaseous atmospheres of these galaxies. 

The main results of our study can be summarized as follows:
\begin{enumerate}
    \item Most TNG galaxies with a stellar mass above a few $10^{9}M_\odot$ host X-ray emitting atmospheres that can easily be detected by {\it Chandra} with a 100 ks exposure (Fig.~\ref{fig:2}). The X-ray morphology of such hot atmospheres can be diverse, with more massive systems hosting more extended and more volume-filling gas than lower-mass objects, and with star-forming galaxies exhibiting biconical features of hot gas extending beyond their cold, star-forming gaseous disks (Fig.~\ref{fig:3}). 
    
    \item After selecting for early-type like galaxies similar to those of available observational datasets taken e.g. from the $\ATLAS$ and MASSIVE surveys, we show that TNG returns $T_{\rm X}-M_{\rm K}$ and $L_{\rm X}-M_{\rm K}$ relations that are consistent with observations (Fig. \ref{fig:5}). This consistency constitutes a non trivial validation of the TNG simulations and of their underlying models for stellar and black hole feedback that are responsible for rearranging and heating the gas within and around galaxies.
    
    \item According to the IllustrisTNG simulations, star-forming and quiescent galaxies exhibit markedly distinct X-ray luminosity vs. K-band magnitude relations. In particular, the TNG simulations predict a clear X-ray luminosity separation between star-forming and quiescent galaxies at $M_{\rm K}\sim -24$, corrsponding to $M_*\sim10^{10.7}M_\odot$, with star-forming galaxies being X-ray {\it brighter} than their quenched counterparts, by up to two orders of magnitudes (Fig. \ref{fig:6}). The difference is more prominent within the central regions ($<R_{\rm e}$) than at larger radii ($5R_{\rm e}$) and it is qualitatively broadly consistent with currently available X-ray data of late and early-type galaxies in the local Universe.
    
    \item On average, the quenched galaxies in IllustrisTNG host gas atmospheres that are hotter but contain significantly less gas than the star-forming galaxies at the same magnitude (Fig. \ref{fig:7}). This indicates that, most likely, the $L_{\rm X}$ diversity between the two populations is driven primarily by gas depletion within quenched galaxies. In other words, quenched galaxies have lower values of $L_{\rm X}$ because they contain less gas than star-forming galaxies, albeit being hotter. This finding is consistent with previous results indicating that gas removal and heating are the primary causes of star formation quenching, at least in TNG.
    
    \item As for the star-formation quenching itself, we show that, according to the TNG simulations, the X-ray luminosity of galactic atmospheres correlates with BH activity and, in particular, the X-ray luminosity dichotomy between star-forming and quiescent galaxies occurs at the same mass scale where the energy injected via SMBH kinetic feedback significantly exceeds the gravitational binding energy of the gas within galaxies (Fig. \ref{fig:8}). This result suggests a direct causal relationship between the SMBH feedback, the physical state of galactic atmospheres, and star-formation.   
\end{enumerate}
The $L_{\rm X}$ dichotomy found in our work has been indirectly addressed in some previous numerical studies (e.g. \citealt{croton.etal.2006,lebrun.etal.2014,choi.etal.2015}) though the discussion in those studies was more about the effects of different SMBH feedback models, e.g. thermal versus kinetic feedback or thermal feedback with various treatments. For instance, \cite{choi.etal.2015} showed that galaxies simulated with thermal feedback are more star-forming (i.e. bluer) and exhibit higher X-ray luminosities than those simulated with mechanical kinetic feedback. However, none of the previous studies addressed the $L_{\rm X}$ diversity problem using a self-consistent model which explains the distinction between the star-forming and quiescent galaxies, as we do in the current study.

To conclude, in this paper we have uncovered an observationally testable, quantitative prediction from the IllustrisTNG simulations. State-of-the-art cosmological simulations of galaxy formation, such as IllustrisTNG (and also EAGLE), support a scenario whereby the quenching of star formation in massive galaxies is caused directly by gas removal from the central regions of galaxies and heating by SMBH feedback. The upcoming all sky survey with {\it eROSITA} will provide the necessary data to perform robust tests for the $L_{\rm X}$ dichotomy between the hot atmospheres of star-forming and quenched galaxies predicted here and to hence further probe the quenching mechanism in the Universe.
\section*{ACKNOWLEDGEMENTS}
We would like to thank the reviewer Benjamin Oppenheimer for constructive comments and suggestions that helped improve the paper. This work was supported by the Lend\"ulet LP2016-11 grant awarded by the Hungarian Academy of Sciences. The authors would like to thank Elad Zinger for useful conversations. The primary TNG simulations were realized with computing time granted by the Gauss Centre for Supercomputing (GCS): TNG50 under GCS Large-Scale Project GCS-DWAR (2016; PIs Nelson/Pillepich) and TNG100 under GCS-ILLU (2014; PI Springel) on the GCS share of the supercomputer Hazel Hen at the High Performance Computing Center Stuttgart (HLRS). GCS is the alliance of the three national supercomputing centres HLRS (Universit{\"a}t Stuttgart), JSC (Forschungszentrum J{\"u}lich), and LRZ (Bayerische Akademie der Wissenschaften), funded by the German Federal Ministry of Education and Research (BMBF) and the German State Ministries for Research of Baden-W{\"u}rttemberg (MWK), Bayern (StMWFK) and Nordrhein-Westfalen (MIWF). Post-processing analyses for this paper were carried out on the Draco and Cobra supercomputers at the Max Planck Computing and Data Facility (MPCDF).

This research has made use of the NASA/IPAC Infrared Science Archive, which is funded by the National Aeronautics and Space Administration and operated by the California Institute of Technology.

\bibliographystyle{mnbst}
\bibliography{ref}
\appendix
\section{X-ray selection and inspection of the mock X-ray analysis}
\label{sec:appA}
\begin{figure*}
    \centering
    \includegraphics[width=0.99\textwidth]{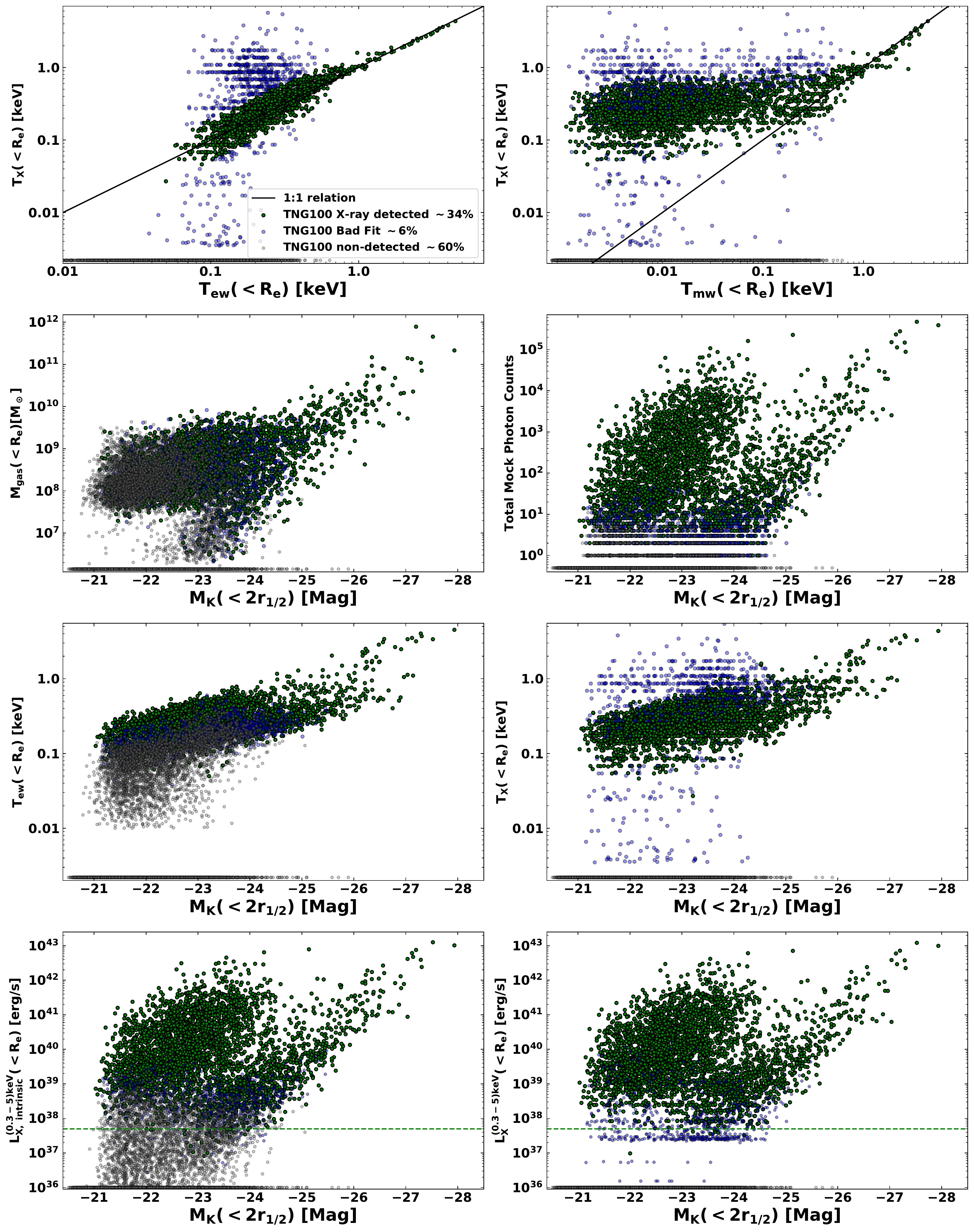}
    \caption{Inspection of the X-ray selection for TNG100 galaxies based on measurements within $R_{\rm e}$. Each relation is shown for three classes of galaxies: ``X-ray detected'', ``bad fit'', and ``non-detected'' (see text for more detail). The horizontal clusters of data points at the bottom of the plots represent the X-ray faint systems with zeros on the y-axes. The dashed lines in the two luminosity-magnitude relations (4th row) denote the value of $5\times10^{37}$ erg/s which approximates the luminosity threshold of the ``X-ray detected'' sample. In the main body of the paper, we only consider for further study the ``X-ray detected'' galaxies from the mock X-ray observations, i.e. the green datapoints.} 
    \label{fig:a0}
\end{figure*}

\begin{figure}
    \includegraphics[width=0.45\textwidth]{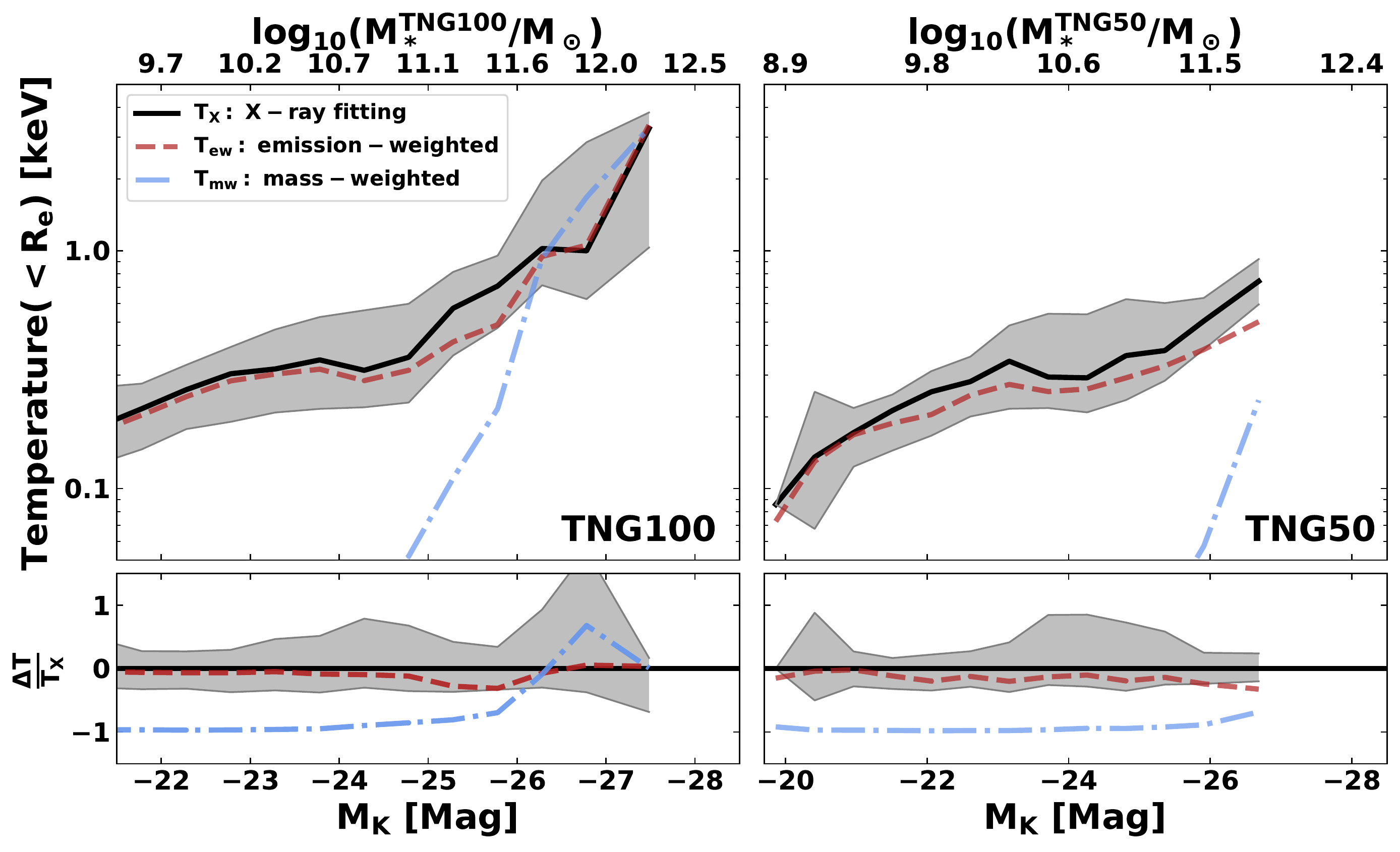}
    \includegraphics[width=0.45\textwidth]{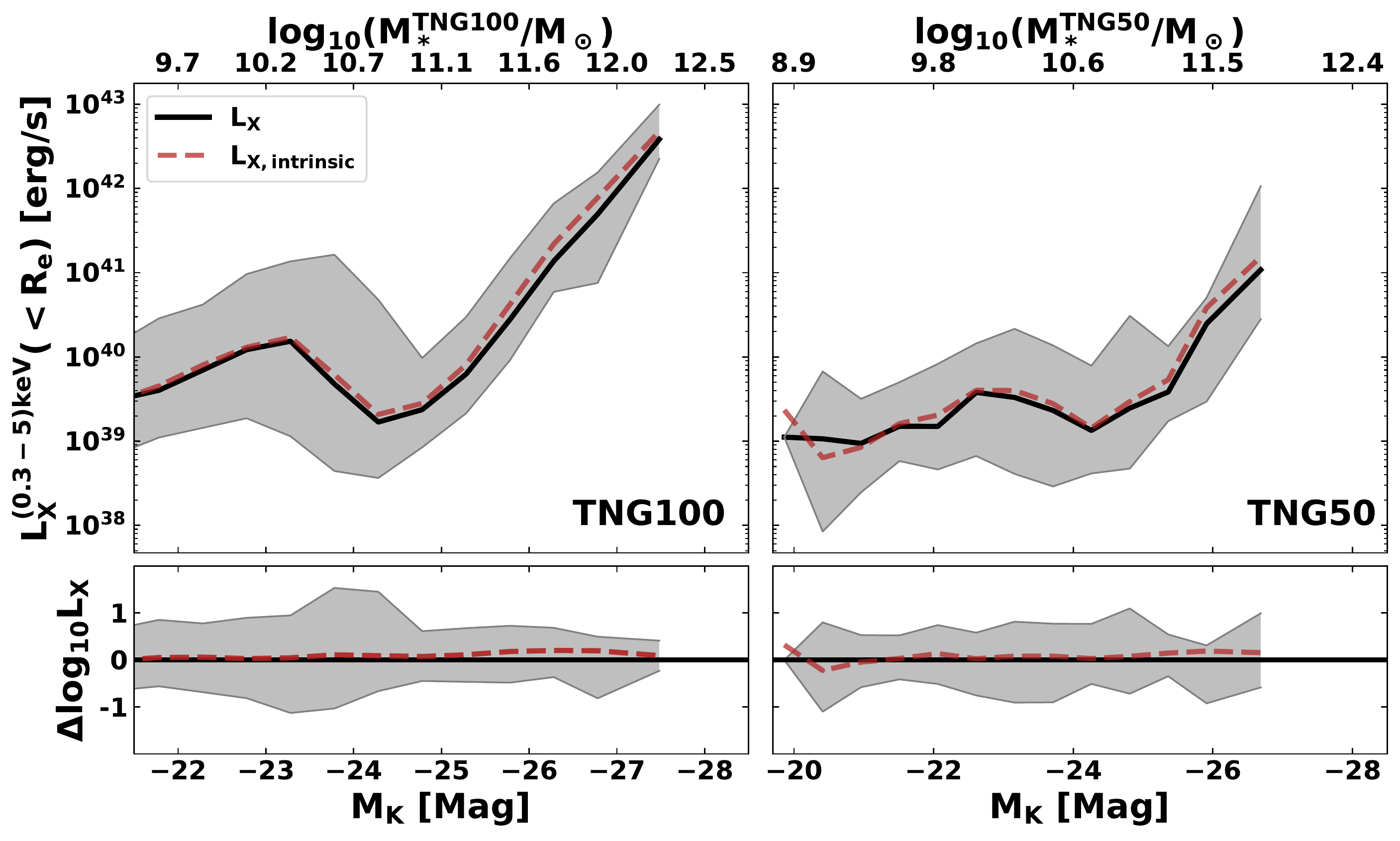}
    \caption{A comparison between the X-ray quantities obtained from the fitting of the mock X-ray spectra and the intrinsic quantities for the ``X-ray detected'' samples of TNG100 ({\it left}) and TNG50 ({\it right}). {\it Top:} comparison between the emission-weighted ($T_{\rm ew}$) and gas mass-weighted ($T_{\rm mw}$) estimators of the gas temperature and the one obtained from fitting mock X-ray spectra ($T_{\rm X}$), as a function of magnitude. The upper panels show the median relations and the lower panels represent the temperature residuals with respect to the best fit values. The shaded areas represent the $1\sigma$ confidence envelope around the median $T_{\rm X}-M_{\rm K}$ relations. {\it Bottom:} similar to the {\it top} plots but for the comparison between the intrinsic X-ray luminosity ($L_{\rm X,intrinsic}$), as given in equation (\ref{eqn2}), and the one inferred from the mock spectral fitting ($L_{\rm X}$).} 
    \label{fig:a1}
\end{figure}
 We compare the X-ray quantities that are obtained from the mock X-ray analysis procedure, described in Section~\ref{sec:mocks}, to the corresponding intrinsic quantities that can be directly computed based on simulated gas cell properties. The comparison is helpful for: i) defining the X-ray selected sample from the simulations, and ii) examining how reliable the assumed fitting model is.
\subsection{X-ray Selection}
Not all galaxies above a certain mass can be characterized by mock X-ray measurements. It depends on the instrument sensitivity (in our case we use {\it Chandra} ACIS-S), the energy band of interest ($[0.3-7]$ keV), the exposure time (we adopt 100 ks), and the physical properties of the gas atmospheres. Galaxies may produce zero or a very low number of photons that are received by {\it Chandra} given the 100 ks exposure time. This may occur because of the limited numerical resolution or for actual physical reasons, e.g. because of the lack of hot gas that could emit photons in the energy range of interest. In the case of a very low photon number, it is not possible to obtain a reliable X-ray temperature by spectral fitting and also $L_{\rm X}$ may not be available.

To gauge the goodness of the results determined by fitting the mock spectra, we compare them with the intrinsic ones (see Section~\ref{sec:mocks}). As it can be appreciated from the top left panel of Fig.~\ref{fig:a0}, the best-fitting temperatures ($T_{\rm X}$) from the X-ray mock analysis appear to follow reasonably well a $1:1$ relation with the emission-weighted estimators ($T_{\rm ew}$), but a number of galaxies deviate strongly from the relation. On the other hand, the best-fitting temperatures tend to overestimate the mass-weighted estimators ($T_{\rm mw}$) in particular at the low-temperature regime ($T_{\rm X}<1$ keV), as shown in the top right panel of Fig.~\ref{fig:a0}. This result could be explained by the fact that in low-mass systems, $T_{\rm mw}$ is mainly determined by gas cells with low temperatures that emit inefficiently in the considered X-ray band.

It should be noted that galaxies producing no or just a handful of photons cannot be fitted: these are indicated as grey data points in Fig.~\ref{fig:a0} (labeled as ``non-detected''). In fact, galaxies with no gas at all (above the simulation resolution limit, see the left panel in the second row of Fig.~\ref{fig:a0}) cannot even be characterized through the intrinsic properties, and therefore are by construction excluded from the analysis of the hot atmospheres. These dominate at low masses ($M_{\rm K} > -24$) and constitute about 60 per cent of galaxies in TNG100. 

We also exclude from the analysis systems with best-fitting mock temperatures $T_{\rm X}$ that lay more than $3\sigma$ off the average $T_{\rm X}-T_{\rm ew}$ relation: blue data points in Fig.~\ref{fig:a0}. Note that the intrinsic scatter of the $T_{\rm X}-T_{\rm ew}$ relation is otherwise rather small: $\sim0.1$ ($\sim0.19$) dex for TNG100 (TNG50) for galaxies with $M_*>3\times10^9M_\odot$ ($M_*>10^8M_\odot$) at $z=0$. The galaxies with ``bad fits'' are a minority but are also characterized by a relatively low numbers of photons: $\lesssim50$, see the right panel in the second row of Fig.~\ref{fig:a0}.

In the third and fourth rows of Fig. \ref{fig:a0}, we inspect the X-ray relations expressed in terms of both the intrinsic quantities ({\it left}), such as $T_{\rm ew}$ and $L_{\rm X,intrinsic}$, and the corresponding quantities obtained from the mock X-ray analysis ({\it right}), $T_{\rm X}$ and $L_{\rm X}$. From the luminosity plots, the ``detected sample'' can be approximately characterised by a threshold in the X-ray luminosity: $L_{\rm X}\gtrsim5\times10^{37}$erg/s.

In conclusion, the analysis is carried out only with those galaxies that, being above a minimum stellar mass (Section~\ref{sec:tng}), are also detected by mock X-ray observations: these are indicated as green data points in Fig.~\ref{fig:a0} and the label ``X-ray detected'' in the main body\footnote{We note that the final sample of ``X-ray detected'' galaxies used in the paper comes from an X-ray selection at $R_{\rm e}$ and $5R_{\rm e}$.}. They constitute about 34 per cent (10 per cent) of the TNG100 (TNG50) galaxies above $M_*>3\times10^9M_\odot$ ($M_*>10^8M_\odot$). 

\subsection{Mock vs. intrinsic quantities}
In Fig. \ref{fig:a1}, we present a quantitative comparison between the mock X-ray fitting results and the intrinsic values for the ``X-ray detected'' samples of TNG100 and TNG50. In the {\it top} plot, we compare the fitted gas temperature with two other temperature estimators: the gas mass-weighted and the emission-weighted, computed according to equation (\ref{eqn3}), as a function of magnitude. Those quantities are averaged temperatures obtained by using different weights, e.g. gas mass and X-ray emission in the $[0.3-5]$ keV range. Compared to the best fit temperatures determined from the mock spectra, the emission-weighted estimator follows well the trend across the considered range of magnitudes. The difference between the two temperatures at a given magnitude in both TNG100 and TNG50 simulations is $<30\%$, well within the intrinsic scatter of the $T_{\rm X}-M_{\rm K}$ relation. On the other hand, the $T_{\rm X}-M_{\rm K}$ is consistent with the $T_{\rm mw}-M_{\rm K}$ relation only for galaxies at the bright end ($M_{\rm K}<-25$ in TNG100). At the faint end, for both the TNG100 and TNG50, the $T_{\rm X}$ significantly overestimates the $T_{\rm mw}$ estimator, as discussed in the previous Section, up to $100\%$, significantly exceeding the intrinsic scatter. The fact that the $T_{\rm X}-M_{\rm K}$ relation follows consistently the $T_{\rm ew}-M_{\rm K}$ relation indicates that the single temperature model that we used for the fitting of the mock X-ray spectra is approximately adequate for studying the temperature relations.

In the {\it bottom} plot of Fig. \ref{fig:a1}, we inspect the X-ray luminosity derived from mock spectral fitting ($L_{\rm X}$) and the intrinsic luminosity ($L_{\rm X,intrinsic}$), that is directly obtained from summing all the gas cell emission as given in equation (\ref{eqn2}), in the $[0.3-5]$ keV range. The fitting-inferred $L_{\rm X}$ is fully compatible with the direct estimation of the total gas cell luminosity within the considered region (i.e. $<R_{\rm e}$), for both simulated datasets. This result ensures that most likely no systematic bias exists in luminosity obtained from the mock X-ray analysis. 
\section{X-ray Emitting Gas in TNG Simulations}
\label{sec:appB}
\begin{figure*}
    \centering
    \includegraphics[width=0.99\textwidth]{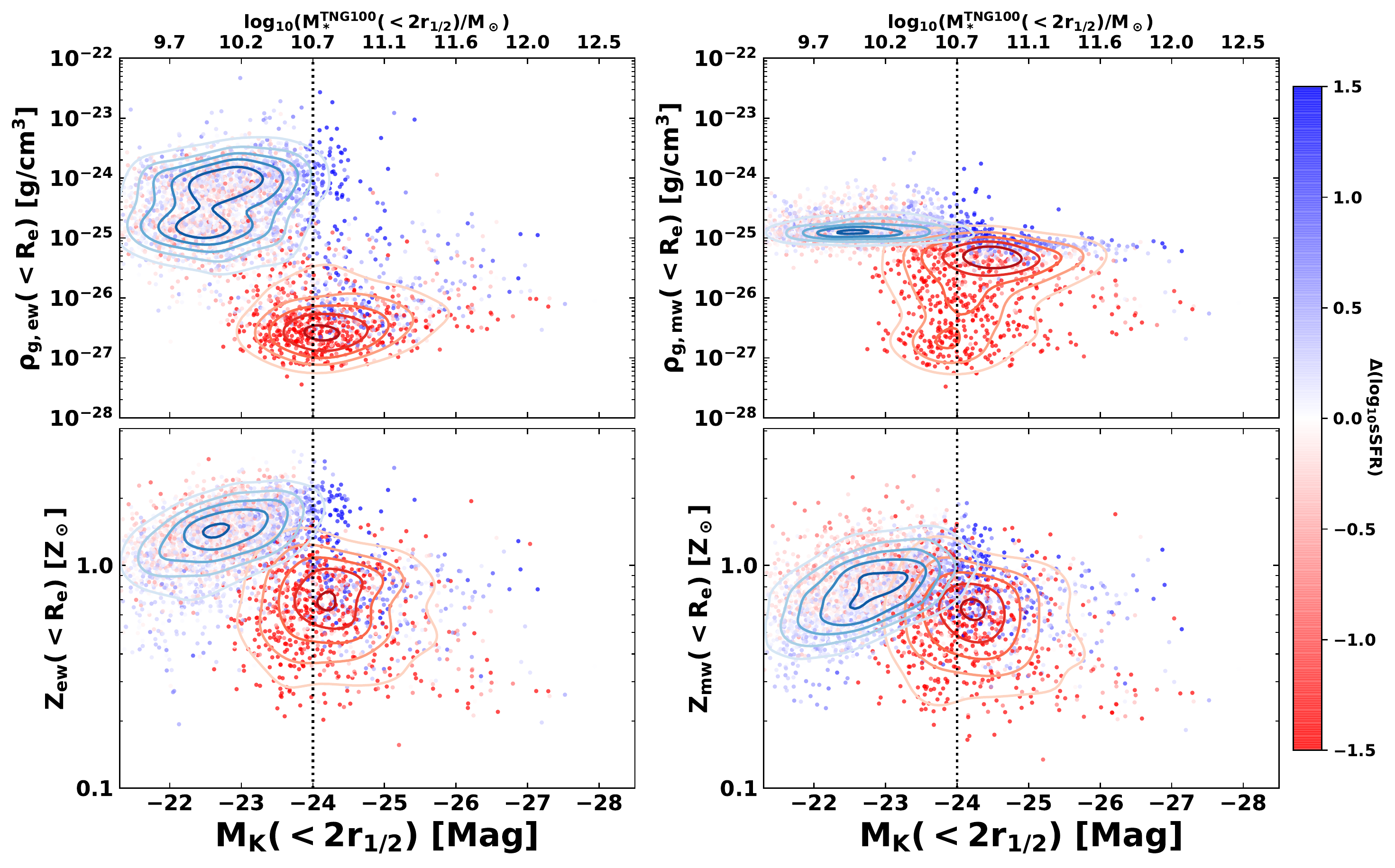}
    \caption{Inspection of the X-ray emitting gas in TNG100. In clock-wise order, emission-weighted gas density, mass-weighted density, mass-weighted metallicity, and emission-weighted metallicity are shown as a function of magnitude. The definitions of the contours and color scheme are identical to Fig. \ref{fig:7}, where the analog investigations for the gas temperature are provided.} 
    \label{fig:b1}
\end{figure*}
In Fig.~\ref{fig:b1} we present the inspection of the properties of X-ray emitting gas in the TNG100 simulation. In addition to the gas temperature, shown in the second row of Fig.~\ref{fig:7}, we further examine the emission-weighted gas density and metallicity in comparison to their mass-weighted estimators. While the latter presumably represents the gas intrinsic properties, the former is mainly determined by the gas component that is responsible for X-ray emission.

Fig.~\ref{fig:b1} shows that the two estimators differ mainly at the low-mass end ($M_{\rm K}>-24$), where $\rho_{\rm g,ew}$ ($Z_{\rm ew}$) is significantly higher than $\rho_{\rm g, mw}$ ($Z_{\rm mw}$), i.e. $\rho_{\rm g,ew}\sim10^{-24}\ {\rm g/cm^3}$ ($Z_{\rm ew}\sim1.5\ Z_\odot$) compared to $\rho_{\rm g,mw}\sim10^{-25}\ {\rm g/cm^3}$ ($Z_{\rm mw}\sim0.7\ Z_\odot$). This result implies that in low-mass galaxies, the X-ray emission mainly comes from condensed and metal-rich gas. For high-mass galaxies, there is no significant difference between the two estimators in both density and metallicity. Furthermore, due to the fact that the emission-weighted gas density and metallicity are biased high, the separation between star-forming and quenched galaxies on the $\rho_{\rm g,ew}-M_{\rm K}$ and $Z_{\rm ew}-M_{\rm K}$ planes appears to be more visible than in the case using the mass-weighted estimators.

Finally, we note that the total soft X-ray emission is dominantly contributed by metals line emission: namely, $\sim84\%-94\%$ for galaxies in the TNG100 sample (see also \citealt{crain.etal.2013}). To evaluate the relative contribution of gas metallicity to the X-ray emission, we carried out an experiment in which we deliberately placed to zero the metallicity value of all selected gas cells and re-calculated $L_{\rm X}$ for the considered sample of TNG100 galaxies and compared the obtained values to the original results (i.e. in the case with non-zero metallicity). Via this test, we have also verified that, even without the contribution from metal emission lines and even though overall the X-ray luminosity is remarkably reduced across the whole range of mass, the $L_X$ dichotomy between star-forming and quenched galaxies is preserved. This confirms that the diversity in the gas mass between the two populations, as discussed in Section \ref{sec:5}, is the main cause of the luminosity dichotomy and not the fact that quiescent galaxies have on average lower gas metallicities than their star-forming counterparts.

\section{Comment on the X-ray Contribution from unresolved ISM phases in TNG Simulations}
\label{sec:appC}
When it comes to estimating the X-ray emission of ISM (e.g. within $R_{\rm e}$), especially in star-forming galaxies, the issue of whether to ignore the contribution of simulated star-forming gas cells becomes a legitimate concern. As mentioned in Section~\ref{sec:mocks}, we note that the contribution of star-forming gas cells to the total X-ray emission is negligible due to their overall low temperatures. In this Section, we further comment on a technical issue regarding the X-ray contribution from the hot component of the unresolved ISM phases.

We recall that the TNG star formation model is based on the subgrid two-phase model proposed by \citealt[][ SH03 hereafter]{springel.hernquist.2003}, with a couple of modifications (see \citealt{pillepich.etal.2018} and references therein): i) the effective pressure predicted by the original model is reduced, through the so-called softer equation of state; and ii) the SN energy is used to power kinetic winds (instead of being dumped into a thermal reservoir of the ISM). Firstly, it should be noted that in the TNG model the SN-driven kinetic, decoupled winds ultimately deposit energy into non-star-forming gas cells, which are already accounted for in our estimation of the X-ray luminosity. Secondly, the hot component of the two-phase model exhibits typically high temperature ($T_{\rm h}>10^{5}$ K, see Fig. 1 in SH03 paper): it hence could in principle emit efficiently in the soft X-ray band (i.e. $[0.3-5]$ keV). Even though formally we could assess the X-ray luminosity contribution from such hot ISM component, we argue that the result would be in practice unphysical due to the fact that this multiphase structure is not resolved in the TNG simulations and it is instead modelled by a simplistic two-phase structure with unrealistic assumptions, e.g. being completely optically thin and with no explicit modelling of the molecular gas (see e.g. \citealt{diemer.etal.2018,stevens.etal.2019} for postprocessing modelling of molecular and atomic gas).

In summary, within the TNG framework we cannot make a sensible estimate of the X-ray emission from the unresolved phases of the ISM. Our practice of excluding star-forming gas cells in estimating the X-ray emission remains the physically preferable and conceptually most robust and best motivated approach for the problem at hand, with the acknowledgement that the quantitative results presented in the main analysis of this paper could constitute lower limits for the X-ray emission from star-forming and quiescent galaxies. Also past analyses for the X-ray and metal ion estimates of simulated galaxies with either the TNG or the EAGLE model have had to compromise in relation to the predictive limitations imposed by the crude treatment of the ISM and have opted for either ignoring at once the star-forming gas \citep{nelson.etal.2018b} or for ignoring gas below $10^5$K or for setting the star-forming gas to $10^4$K, the typical temperature of the warm-neutral ISM \citep[e.g.][]{lebrun.etal.2014,rahmati.etal.2016,wijers.etal.2019}.

\end{document}